%% file: main.tex
\documentclass[
    twocolumn,
    10pt,
    superscriptaddress,
    aps,
    pra,
    nofootinbib,
]{revtex4-2}
\usepackage[T1]{fontenc}
\usepackage[dvipsnames]{xcolor}
\usepackage{float}
\usepackage{inputenc}
\usepackage{amsthm}
\usepackage{thmtools}
\usepackage{mathtools}
\usepackage{physics}
\usepackage{afterpage}
\usepackage{bm} 
\usepackage{bbm} 
\usepackage{braket}
\usepackage{dsfont}
\usepackage{array}
\newcolumntype{M}[1]{>{\centering\arraybackslash}p{#1}}
\newenvironment{pMatrix}[1]
  {\left(\begin{array}{*{#1}{M{0.37cm}}}} 
  {\end{array}\right)}
\usepackage{tabularx}
\newcolumntype{C}{>{\centering\arraybackslash}X}
\usepackage{amsmath}
\usepackage{amssymb}
\usepackage{tikz}
\usetikzlibrary{arrows, positioning}
\usepackage{tikz-cd}
\usepackage{ulem}
\usepackage[linesnumbered,ruled]{algorithm2e}
\SetArgSty{textnormal}

\usepackage{etoolbox}
\apptocmd{\sloppy}{\hbadness 10000\relax}{}{}

\usepackage{silence}
\definecolor{QK_plot}{HTML}{1188FF}
\definecolor{QK_extr}{HTML}{0235AA}
\definecolor{TPD_plot}{HTML}{FF1188}
\definecolor{TPD_extr}{HTML}{AA0235}
\definecolor{HY_plot}{HTML}{11FF88}
\definecolor{HY_extr}{HTML}{02AA35}

        \DeclareMathOperator{\id}{id} 
        \DeclareMathOperator{\Mat}{Mat} 
        
        \newcommand*{\N}{\mathbb{N}}

        \newcommand*{\C}{\mathbb{C}}
        \newcommand*{\hil}{\mathcal{H}}
        \newcommand*{\defdot}{\,\cdot\,} 
        \newcommand*{\defcolon}{\,:\,} 
        \newcommand\scriptin{\raisebox{0.15ex}{$\scriptscriptstyle\in$}} 
        
        \newcommand*{\one}{\mathds{1}} 
        \newcommand*{\lo}{\mathcal{L}} 
        \newcommand*{\quater}{\text{Q}} 
        \newcommand*{\rstring}{\mathbf{r}} 
        \newcommand*{\sstring}{\mathbf{s}} 
        \newcommand*{\tstring}{\mathbf{t}} 
        \newcommand*{\cmw}{\Omega} 
        \newcommand*{\bigo}{\mathcal{O}} 
        \newcommand*{\pbasis}{\mathcal{P}} 
        \newcommand*{\cbasis}{\mathcal{C}} 

\usepackage[colorlinks]{hyperref}
\hypersetup{
    colorlinks  = true,
    citecolor   = PineGreen,
    linkcolor   = MidnightBlue,
    urlcolor    = PineGreen
}

\makeatletter
\@ifclasswith{revtex4-2}{linenumbers}{
    \NewCommandCopy\oldequation\equation
    \NewCommandCopy\endoldequation\endequation
    \renewenvironment{equation}{\linenomath\oldequation}
    {\endoldequation\endlinenomath}

    \NewCommandCopy\oldalign\align
    \NewCommandCopy\endoldalign\endalign
    \renewenvironment{align}{\linenomath\oldalign}
    {\endoldalign\endlinenomath}
}{}
\makeatother

\begin{document}

\title{Fast generation of Pauli transfer matrices utilizing tensor product structure}

\begin{abstract}
    Analysis of quantum processes, especially in the context of noise, errors, and decoherence is essential for the improvement of quantum devices.
    An intuitive representation of those processes modeled by quantum channels are Pauli transfer matrices.
    They display the action of a linear map in the $n$-qubit Pauli basis in a way, that is more intuitive, since Pauli strings are more tangible objects than the standard basis matrices.
    We set out to investigate classical algorithms that convert the various representations into Pauli transfer matrices.
    We propose new algorithms that make explicit use of the tensor product structure of the Pauli basis.
    They convert a quantum channel in a given representation (Chi or process matrix, Choi matrix, superoperator, or Kraus operators) to the corresponding Pauli transfer matrix.
    Moreover, the underlying principle can also be used to calculate the Pauli transfer matrix of other linear operations over $n$-qubit matrices such as left-, right-, and sandwich multiplication as well as forming the (anti-)commutator with a given operator.
    Finally, we investigate the runtime of these algorithms, derive their asymptotic scaling and demonstrate improved performance using instances with up to seven qubits.
\end{abstract}

\author{Lukas Hantzko}
\affiliation{Institut f\"ur Theoretische Physik, Leibniz Universit\"at Hannover, Germany}
\thanks{Submitted to: \textit{Phys. Scr.}}
\author{Lennart Binkowski}
\email{lennart.binkowski@itp.uni-hannover.de}
\affiliation{Institut f\"ur Theoretische Physik, Leibniz Universit\"at Hannover, Germany}
\author{Sabhyata Gupta}
\affiliation{Institut f\"ur Theoretische Physik, Leibniz Universit\"at Hannover, Germany}

\maketitle

\section{\label{section:Introduction}Introduction}

Characterization of unknown quantum states and quantum processes is a critical task in quantum computing, quantum communication, and quantum information processing.
Pauli Transfer Matrix (PTM)-based representation provides a versatile and natural formalism for representing quantum processes due to its inherent alignment with the structure of quantum systems and  applicability across a broad spectrum of quantum computing tasks.
Its utility extends to various fields such as quantum process tomography (see \cite{Roncallo2023PauliTransferMatrixDirectReconstructionChannelCharacterizationWithoutFullProcessTomography} for an application involving Pauli measurements) classical simulations of quantum circuits (e.g.\ in \cite{Rall2019SimulationOfQubitQuantumCircuitsViaPauliPropagation} PTMs are used to identify quantum channels as simulable), as well as randomized benchmarking~\cite{Corcoles2013ProcessVerificationOfTwoQubitQuantumGatesByRandomizedBenchmarking} and compiling~\cite{Wallman2016NoiseTailoringForScalableQuantumComputationViaRandomizedCompiling} where PTMs serve as the standard tool to analyze the effect of \emph{Pauli twirling}.

Quantum channels modeling noisy processes and various errors models can be systematically represented with PTMs, as several such channels are most naturally represented in the Pauli basis (see, e.g., \cite{Kervinen2024ExtendedQuantumProcessTomographyOfLogicalOperationsOnAnEncodedBosionicQubit} for a thorough application).
Additionally, PTMs offer a way to represent quantum channels, which is easier to visualize than other representations, e.g.\ Kraus operators, Choi or Chi matrices or the conventional superoperator representation (see \cite{Wood2015TensorNetworksAndGraphicalCalculusForOpenQuantumSystems} for an excellent overview).
PTMs are also widely used in the analysis of error correcting codes (in \cite{Marques2022LogicalQubitOperationsInAnErrorDetectingSurfaceCode} PTMs are specifically used to quantify the performance of fault-tolerant, logical gates), error mitigation (see \cite{Cai2020MitigatingCoherentNoiseUsingPauliConjugation} for the application of Pauli conjugation to mitigate coherent noise) and Hamiltonian learning (\cite{Caro2024LearningQuantumProcessesAndHamiltoniansViaThePauliTransferMatrix} discusses learning PTMs of general quantum processes and extends this technique to Hamiltonian learning).
This simply arises from the fact that the Pauli basis is a frequently used basis.
This can be seen e.g.\ in measurement-based quantum computation~\cite{Raussendorf2003MeasurementBasedQuantumComputationOnClusterStates} or in quantum error correction using stabilizers.
Therefore the PTM is compatible also for experimental investigation of quantum channels.

In this work, we primarily focus on the task of converting various representations of quantum channels as well as special linear maps into the PTM representation as a numerical task on a classical computer.
We believe that enhancing computational efficiency in this conversion process will facilitate the analysis and classical simulation of quantum channels involving a larger number of qubits.

Existing conversion methods such as the ones within Forest~\cite{Gulshen2019ForestBenchmarkingQCVVUsingPyQuil} and IBM's Qiskit~\cite{JavadiAbhari2024QuantumComputingWithQiskit}, rely on first converting any given channel to the superoperator representation and from there to the desired one.
They both have rather straightforward implementations of the conversions described in \cite{Wood2015TensorNetworksAndGraphicalCalculusForOpenQuantumSystems}.
Their conversion from the superoperator representation involves the computation of the change-of-basis matrix and matrix multiplication.
In contrast, we present a set of algorithms based on our recently proposed Tensorized Pauli Decomposition (\texttt{TPD}) algorithm~\cite{Hantzko2024TensorizedPauliDecompositionAlgorithm}, which allow direct conversion between any channel representation and the PTM.
As we show in our complexity analysis, taking such a shortcut significantly reduces the runtime of the conversion process.

Furthermore, we derive algorithms with similar design principles for some special superoperators (which are not necessarily quantum channels).
Namely, we introduce routines that translate the left, right, and sandwich multiplication with multi-qubit operators as well as forming the (anti-)commutator with such an operator into the PTM representation, respectively.
As such terms arise frequently in application, we believe that these special case-algorithms will have an impact on practical computations comparable to the one of our algorithms for translating various channel representations into the PTM.

\begin{figure*}[!ht]
    \centering
    \input{Trafo.tex}
    \caption{Transformations between different channel representations.
    Qiskit provides algorithms for most transformations as indicated by the arrows.
    Typically, these involve two steps: translate a given representation into the canonical one and then translate further to the target representation.
    In this paper, we explicitly construct algorithms to perform direct conversions from all representations to PTM that may also be applied to further conversion such as between Choi and Chi matrices ({\color{JungleGreen} green} lines);
    Except for the \texttt{Kraus-PTM} algorithm, our methods are experimentally faster on instances up to $7$ qubits and admit a favorable asymptotic runtime.
    In particular, the \texttt{Can-PTM} algorithm achieves a speed-up for translating from the canonical representation to PTM, therefore one can simply substitute this algorithm for Qiskit's version to also achieve a speed-up for the conversion from Kraus representation to PTM.}
    \label{figure:DiagramOfRepresentations}
\end{figure*}

The rest of the article is organized as follows:
We begin \autoref{section:Preliminaries} with some mathematical foundations regarding representations of linear maps and the properties of translating between different representations.
We fill these abstract notions with life by subsequently discussing concrete representations of quantum channels.
We conclude the section with a different take on the task of calculating Pauli decompositions of matrices, offering valuable insights into the \texttt{TPD} and its inverse operation.
In \autoref{section:Methods}, we introduce various algorithms for calculating the PTM of the aforementioned special cases of superoperators as well as for the common channel representations (canonical, Choi, Chi, and Kraus).
These include more algebraic relations using the cumulative matrix weights, as they can be used to define iterative basis change or the tensor product structure for the Choi-matrix.
Lastly, in \autoref{section:Results}, we analyze the proposed algorithms' complexities to demonstrate that they can asymptotically compete with the currently used implementations within the Forest benchmarking library and  Qiskit, while offering practical speed-ups by avoiding many intermediate steps.

\section{\label{section:Preliminaries}Preliminaries}

\subsection{\label{subsection:QuantumChannelsAndTheirRepresentations}Quantum channels and their representations}

Given two quantum systems A and B, described by Hilbert spaces $\hil_{\text{A}}$ and $\hil_{\text{B}}$, a quantum channel (in the Schrödinger picture) from system A to system B is a completely positive, trace-preserving (CPTP) superoperator $\mathcal{E} : \lo(\hil_{\text{A}}) \rightarrow \lo(\hil_{\text{B}})$.\footnote{
    Conversely, a quantum channel in the Heisenberg picture is required to be unital instead of trace-preserving.
}
Complete positivity of $\mathcal{E}$ means that for all $d \in \N$ the map $\id_{\Mat(d)} \otimes\, \mathcal{E}$ maps positive elements in $\Mat(d) \otimes \hil_{\text{A}}$ to positive elements in $\Mat(d) \otimes \hil_{\text{B}}$.\footnote{
    The CPTP property ensures that $\mathcal{E}$ and the induced mappings $\id_{\Mat(d)} \otimes\, \mathcal{E}$ map density matrices over any composite system containing A to density matrices over any composite system containing B.
}
In the following, we may restrict ourselves to the concrete case where both A and B are $n$-qubit systems, i.e.\ $\hil_{\text{A}} \cong \hil_{\text{B}} \cong \C^{2^{n}}$, and hence $\lo(\hil_{\text{A}}) \cong \lo(\hil_{\text{B}}) \cong \Mat(2^{n})$.

In order to treat different representations of quantum channels properly, it is important to cautiously distinguish between abstract linear operators and their possible matrix representations.
Although this distinction is very important on the level of superoperators, on the level of density matrices/operators this introduces some conceptual overhead that may not always be justified from an application-focused point of view.
As a compromise we make as many underlying objects as explicit as possible to eliminate ambiguity.
Namely, we set $\hil_{\text{A}} = \hil_{\text{B}} = \C^{2^{n}}$.
Furthermore, we set $\lo(\hil_{\text{A}}) = \lo(\hil_{\text{B}}) = \Mat(2^{n})$ and introduce its \emph{canonical basis} 
\begin{align}\label{equation:CanonicalBasis}
    \cbasis \coloneqq \big\{E_{k, \ell} \coloneqq (\delta_{i, k} \delta_{j, \ell})_{1 \leq i, j \leq 2^{n}} : k, \ell = 1, \ldots, 2^{n}\big\},
\end{align}
which readily is an orthogonal basis of $\Mat(2^{n})$.
Note however that, in our convention for the Frobenius inner product, $\cbasis$ is not normalized, i.e.\ is not an \emph{orthonormal basis} (ONB).

A second important matrix basis of $\Mat(2^{n})$ consists of tensor products of Pauli matrices:
The Pauli matrices $\sigma^{\{0, 1, 2, 3\}} \coloneqq \{I, X, Y, Z\}$ constitute a set of hermitian, involutory, and unitary $2 \times 2$ matrices.
Defining the index set $\quater \coloneqq \{0, 1, 2, 3\}$, a \emph{Pauli string} of length $n$ is associated with a quaternary string $\tstring \in \quater^{n}$ and given by
\begin{align}\label{equation:PauliString}
    \sigma^{\tstring} \coloneqq \sigma^{\text{t}_{1}} \otimes \sigma^{\text{t}_{2}} \otimes \cdots \otimes \sigma^{\text{t}_{n - 1}} \otimes \sigma^{\text{t}_{n}}.
\end{align}
Pauli strings are again hermitian, involutory, and unitary matrices, and the set of all length-$n$ Pauli strings, the \emph{Pauli basis}
\begin{align}\label{equation:PauliBasis}
    \pbasis \coloneqq \{\sigma^{\tstring} \defcolon \tstring \in \quater^{n}\}
\end{align}
forms an ONB of $\Mat(2^{n})$.

Quantum channels can be represented in many different forms, each of which can be converted into the others.
One common possibility of representing a given channel $\mathcal{E} \in \lo(\Mat(2^{n}))$ is to choose an orthogonal basis $\mathcal{B}$ of $\Mat(2^{n})$ as well as an \emph{explicit} ordering of its basis elements $\{B_{i}\}_{i = 1}^{4^{n}}$ and to collect the entries $\langle B_{i}, \mathcal{E}(B_{j})\rangle$, $i, j = 1, \ldots, 4^{n}$ in a $4^{n} \times 4^{n}$ matrix.
To be more precise, this establishes an algebra-isomorphism
\begin{align}\label{equation:BasisRepresentation}
\begin{split}
    [\defdot]_{\mathcal{B}} : \lo(\Mat(2^{n})) &\rightarrow \Mat(4^{n}) \\
    \mathcal{E} &\mapsto (\langle B_{i}, \mathcal{E}(B_{j})\rangle)_{1 \leq i, j \leq 4^{n}}.
\end{split}
\end{align}

A natural choice for such a matrix basis is the canonical basis \eqref{equation:CanonicalBasis} with its \emph{standard ordering} 
\begin{align}
    E_{i} \coloneqq E_{i(k, \ell)} \coloneqq E_{\ell + 2^{n} (k - 1)} \coloneqq E_{k, \ell}.
\end{align}
We may refer to this representation also as the \emph{canonical} one and denote it by Can$(\mathcal{E})$.
A second candidate is the aforementioned Pauli basis \eqref{equation:PauliBasis} and the resulting matrix -- the \emph{Pauli transfer matrix} (PTM) of $\mathcal{E}$ -- holds the entries PTM$(\mathcal{E})_{\sstring, \tstring} = \langle \sigma^{\sstring}, \mathcal{E}(\sigma^{\tstring})\rangle$ in lexicographic order in both $\sstring, \tstring \in \quater^{n}$.

Taking $\mathcal{B} \in \{\cbasis, \pbasis\}$ further establishes that \eqref{equation:BasisRepresentation} factorizes in a tensor product of linear maps for the tensor decompositions
$\lo(\Mat(2^{n})) \cong \lo(\Mat(2^{n_{1}})) \otimes \lo(\Mat(2^{n_{2}}))$ and $\Mat(4^{n}) \cong \Mat(4^{n_{1}}) \otimes \Mat(4^{n_{2}})$ for $n_{1}, n_{2} \geq 1$ such that $n_{1} + n_{2} = n$:
Both the canonical basis and the Pauli basis of $\Mat(2^{n})$ can be written as the set of all the pure tensors formed from the canonical bases of $\Mat(2^{n_{1}})$ and $\Mat(2^{n_{2}})$, and the Pauli bases of $\Mat(2^{n_{1}})$ and $\Mat(2^{n_{2}})$, respectively;
as the order of basis elements matters, we have to explicitly state an enumeration rule in both cases (compare the standard ordering of the canonical basis).
Namely, for the canonical basis we have that
\begin{align}\label{equation:CanonicalBasisPureTensor}
    E_{i, j}^{(1)} \otimes E_{k, \ell}^{(2)} = E_{2^{n_{2}} (i - 1) + k, 2^{n_{2}} (j - 1) + \ell}^{\vphantom{(0)}}
\end{align}
and for the Pauli basis it holds that
\begin{align}\label{equation:PauliBasisPureTensor}
    \sigma^{\rstring} \otimes \sigma^{\sstring} = \sigma^{\rstring \sstring}, \text{ where } \rstring \in \quater^{n_{1}} \text{ and } \sstring \in \quater^{n_{2}}.
\end{align}
The difference between the two bases lies in the ordering. While the Pauli basis has the lexicographic order of $\quater^{n}$, the canonical basis does not.
If a relation such as \eqref{equation:CanonicalBasisPureTensor} or \eqref{equation:PauliBasisPureTensor} is satisfied and moreover the order is lexicographic that is $B^{(n)}_{4^{n - 1} (i - 1) + j} = B^{(1)}_{i} \otimes B^{(n-1)}_{j}$, the basis is called \emph{tensorial}.
Given two quantum channels $\mathcal{E}_{i} \in \lo(\Mat(2^{n_{i}}))$, $i = 1, 2$ and such a tensorial ONB of pure tensors, the matrix representation of $\mathcal{E}_{1} \otimes \mathcal{E}_{2}$ indeed factorizes into the Kronecker product of the representations of $\mathcal{E}_{1}$ and $\mathcal{E}_{2}$.
Therefore, we have just established that
\begin{align}
    \text{PTM}(\mathcal{E}_{1} \otimes \mathcal{E}_{2}) &= \text{PTM}(\mathcal{E}_{1}) \otimes \text{PTM}(\mathcal{E}_{2}). \label{equation:TensoralityPTM}
\end{align}

Another matrix representation w.r.t.\ the canonical basis is given by the \emph{Choi matrix} with
\begin{align}\label{equation:ChoiIsomorphism}
\begin{split}
    \text{Choi} : \lo(\Mat(2^{n})) &\rightarrow \Mat(4^{n}) \\
    \mathcal{E} &\mapsto \sum_{k, \ell = 1}^{2^{n}} E_{k, \ell} \otimes \mathcal{E}(E_{k, \ell}),
\end{split}
\end{align}
where \eqref{equation:ChoiIsomorphism} is a basis-dependent version of the more general Choi-Jamio\l{}kowski isomorphism, after \citet{Choi1975CompletelyPositiveLinearMapsOnComplexMatrices} and \citet{Jamiolkowski1972LinearTransformationsWhichPreserveTraceAndPositiveSemidefinitenessOfOperators}, (see also \cite{Frembs2024VariationsOnTheChoiJamiolkowskiIsomorphism} for further discussions) which establishes that the complete positivity of $\mathcal{E}$ is equivalent to $\text{Choi}({\mathcal{E}})$ being a positive matrix.\footnote{Normalizing the Choi matrix to unit trace yields the Choi state.}
Note that $\Mat(4^{n})$ has to be equipped with a different multiplication in order to establish multiplicativity of \eqref{equation:ChoiIsomorphism}, turning it into an algebra-isomorphism:
the \emph{link product}
\begin{align}\label{equation:LinkProduct}
    F \star G \coloneqq \tr_{2}\Big[\big(\one_{\Mat(2^{n})} \otimes F\big) \big(G^{\text{T}_{2}} \otimes \one_{\Mat(2^{n})}\big)\Big].
\end{align}
Here, the matrices $F, G \in \Mat(4^{n})$ are interpreted as elements in $\Mat(2^{n}) \otimes \Mat(2^{n})$, ``$\tr_{2}$'' denotes the partial trace over the second tensor factor, and ``$\text{T}_{2}$'' denotes the partial transpose in the second tensor factor w.r.t.\ the canonical basis.

The Choi matrix is an example of a \emph{Chi-matrix} $\text{Chi}(\mathcal{E}) \in \Mat(4^{n})$ \cite{Wood2015TensorNetworksAndGraphicalCalculusForOpenQuantumSystems} with the canonical basis as reference basis.
In the following, however, we reserve the name Chi for the Pauli basis as reference basis.
That representation has the following connection to the channel $\mathcal{E}$:
\begin{align}\label{equation:ChiMatrix}
    \mathcal{E}(\rho) = \sum_{\sstring, \tstring \scriptin \quater^{n}} \text{Chi}(\mathcal{E})_{\sstring, \tstring} \sigma^{\sstring} \rho \sigma^{\tstring}.
\end{align}
The relation between Choi and Chi matrix is given by a change of basis from the canonical to the Pauli basis.

Another alternative is given by the \emph{Kraus representation}:
The proof of the Choi-Jamio\l{}kowski theorem further entails that $\mathcal{E}$ has the form
\begin{align}\label{equation:KrausRepresentation}
    \mathcal{E}(\rho) = \sum_{i} K_{i}^{\vphantom{\dagger}} \rho K_{i}^{\dagger}
\end{align}
with at most $4^{n}$ \emph{Kraus operators} $K_{i} \in \Mat(2^{n})$ such that $\sum_{i} K_{i}^{\dagger} K_{i}^{\vphantom{\dagger}} = \one$.
Note, however, that the Kraus operators are not uniquely determined, not even their number.
The minimum number of Kraus operators such that $\mathcal{E}$ can be written in the form \eqref{equation:KrausRepresentation} is called the \emph{Kraus rank} of $\mathcal{E}$.
More generally, $\mathcal{E}$ may be expressed as
\begin{align}\label{equation:GeneralizedKrausRepresentation}
    \mathcal{E}(\rho) = \sum_{i} K_{i}^{\vphantom{\dagger}} \rho L_{i}^{\dagger}
\end{align}
with \emph{generalized Kraus operators} $K_{i}, L_{i} \in \Mat(2^{n})$.

In this paper, we address the objective of translating the canonical, the Choi, the Chi, and the (generalized) Kraus representation into the PTM, respectively.
Furthermore, we will investigate superoperators that have special forms (such as taking the commutator with a given Hamiltonian) and derive special-case algorithms for constructing their PTMs.

\subsection{\label{subsection:PauliDecompositionAsBasisChange}Pauli decomposition as basis change}

\begin{figure*}[!th]
    \begin{minipage}[t]{0.47\linewidth}
        \begin{algorithm}[H]\label{algorithm:LRPTM}
            \caption{\texttt{L/R-PTM}($A$)}
            \eIf{$\dim(A) = 1 \times 1$} {
                \Return{$A$}\;
            } {
                $B \coloneqq 0$\;
                \For{$\text{t} \in \quater$}{
                    compute $\cmw^{\vphantom{(1)}}_{\text{t} \bm{\ast}}$ according to \eqref{equation:CumulativeWeightsDefinition}\;
                    \If{$\cmw^{\vphantom{(1)}}_{\text{t} \bm{\ast}} \neq 0$} {
                        $B \mathrel{+}= \epsilon\texttt{-L/R-PTM}[\text{t}] \otimes \texttt{L/R-PTM}(\cmw^{\vphantom{(1)}}_{\text{t} \bm{\ast}})$\;
                        \tcp{$\epsilon\texttt{-L/R-PTM}[\text{t}]$: \hspace*{-12pt} elementary PTM t}
                        \tcp{for left/right multiplication}
                    }
                }
                \Return{$B$}\;
            }
        \end{algorithm}
    \end{minipage}\hfill
    \begin{minipage}[t]{0.47\linewidth}
        \begin{algorithm}[H]\label{algorithm:MTM}
            \caption{\texttt{M-PTM}($A_{1}, A_{2}$)}
            \eIf{$\dim(A_{1}) = \dim(A_{2}) = 1 \times 1$} {
                \Return{$A_{1} A_{2}$}\;
            } {
                $B \coloneqq 0$\;        
                \For{$\text{t}, \text{u} \in \quater$}{
                    compute $\cmw^{(1)}_{\text{t} \bm{\ast}}$ and $\cmw^{(2)}_{\vphantom{\text{t}}\text{u} \bm{\ast}}$ according to \eqref{equation:CumulativeWeightsDefinition}\;
                    \If{$\cmw^{(1)}_{\text{t} \bm{\ast}} \neq 0$ and $\cmw^{(2)}_{\vphantom{\text{t}}\text{u} \bm{\ast}} \neq 0$} {
                        $B \mathrel{+}= \epsilon\texttt{-M-PTM}[\text{t}][\text{u}] \otimes \texttt{M-PTM}(\cmw_{\text{t} \bm{\ast}}^{(1)}, \cmw_{\vphantom{\text{t}}\text{u} \bm{\ast}}^{(2)})$\;
                        \tcp{$\epsilon\texttt{-M-PTM}[\text{t}][\text{u}]$: \hspace*{-12pt} elementary PTM $(\text{t}, \text{u})$}
                        \tcp{for sandwich multiplication}
                    }
                }
            }
            \Return{$B$}\;
        \end{algorithm}
    \end{minipage}
\caption{
    \texttt{PTM}-algorithms for single-sided and sandwich-multiplication.
    The left/right multiplication algorithm inputs a single matrix $2^{n} \times 2^{n}$-matrix  $A$.
    If $n = 0$, i.e.\ $A$ already is a scalar, it is simply returned.
    Otherwise, $B$ is initialized as the zero matrix of dimension $2^{n} \times 2^{n}$. Subsequently, $A$'s CMWs are calculated via matrix slicing and, for each nonzero CMW $\cmw_{\text{t} \bm{\ast}}$, the tensor product of the $t$-th element in the lookup table for elementary PTMs for left/right multiplication $\epsilon$\texttt{-L/R-PTM} with the result of \texttt{L/R-PTM}($\cmw_{\text{t} \bm{\ast}}$) -- following \eqref{equation:LeftMultiplication} and \eqref{equation:RightMultiplication} -- is added to $B$.
    Finally, $B$ is returned.
    The algorithm for the sandwich multiplication has a similar structure, but inputs two $2^{n} \times 2^{n}$-matrices for the left and right multiplicand, respectively.
    If $n = 0$, their scalar-valued product is returned.
    Otherwise, for all nonzero CMWs $\cmw^{(1)}_{\text{t} \bm{\ast}}$ and $\cmw^{(2)}_{\vphantom{\text{t}}\text{u} \bm{\ast}}$, all tensor products of the $(\text{t}, \text{u})$-th element in the lookup table for elementary PTMs for sandwich multiplication $\epsilon$\texttt{-M-PTM} with the result of \texttt{M-PTM}($\cmw^{(1)}_{\text{t} \bm{\ast}}, \cmw^{(2)}_{\vphantom{\text{t}}\text{u} \bm{\ast}}$) -- following \eqref{equation:SandwichMultiplication} -- are added to the as zero initialized result matrix, which is subsequently returned.
}
\label{figure:MultiplicationAlgorithms}
\end{figure*}

For a given matrix $A = \sum_{k, \ell} A_{k \ell} E_{k, \ell} \in \Mat(2^{n})$ the objective of calculating its Pauli decomposition
\begin{align}\label{equation:PauliDecomposition}
    A = \sum_{\tstring \scriptin \quater^{n}} \omega_{\tstring} \sigma^{\tstring},\quad \omega_{\tstring} = \langle \sigma^{\tstring}, A\rangle
\end{align}
is nothing but conducting a basis change from the canonical to the Pauli basis.
The \emph{transition matrix} $B$ of dimensions $4^{n} \times 4^{n}$ is implicitly defined via the relation between both bases' elements:
\begin{align}
    \sigma^{\tstring}\hspace*{-0.5pt} =\hspace*{-0.5pt} \sum_{j = 1}^{4^{n}} B_{j \tstring} E_{j},\  \text{where } j = j(k, \ell) = k + 2^{n} (\ell - 1)
\end{align}
indexes the canonical basis in standard ordering.
The components of $A$ in both bases are therefore related via
\begin{equation}
    A = \sum_{j = 1}^{4^{n}} A_{i} E_{j} = \sum_{\tstring \scriptin \quater^{n}} \omega_{\tstring} \sigma^{\tstring} = \sum_{\tstring \scriptin \quater^{n}} \omega_{\tstring} \sum_{j = 1}^{4^{n}} B_{j \tstring} E_{j},
\end{equation}
which means, that the components change with the basis as 
\begin{align}
   A_{j} = \sum_{\tstring \scriptin \quater^{n}} B_{j\tstring} \omega_{\tstring} \quad \text{and}\quad \omega_{\tstring} = \sum_{j = 1}^{4^{n}} \big(B^{-1}\big)_{\tstring j} A_{j}^{\vphantom{-1}}.
\end{align}

The chosen orderings of the canonical and the Pauli basis imply that the basis change matrix $B$ (and its inverse) can be written as an $n$-fold tensor product of an elementary, $4 \times 4$-basis change matrix
\begin{align}
    B = \begin{pmatrix}
        1 & 0 & 0 & 1 \\
        0 & 1 & -i & 0 \\
        0 & 1 & i & 0 \\
        1 & 0 & 0 & -1
    \end{pmatrix},\ 
    B^{-1} = \frac{1}{2} \begin{pmatrix}
        1 & 0 & 0 & 1 \\
        0 & 1 & 1 & 0 \\
        0 & i & -i & 0 \\
        1 & 0 & 0 & -1
    \end{pmatrix}.
\end{align}
Note that $B$ is not unitary due to the fact that the canonical and Pauli basis are not normalized w.r.t.\ the same inner product.

The tensor product structure of the basis change matrix is exactly what is exploited in our recently proposed \emph{Tensorized Pauli Decomposition} (\texttt{TPD}) algorithm~\cite{Hantzko2024TensorizedPauliDecompositionAlgorithm}.
Here, a given matrix $A \in \Mat(2^{n})$ is partitioned into four blocks of dimension of $2^{n - 1} \times 2^{n - 1}$ such that
\begin{align}
    A = \begin{pmatrix}
        A_{1 1} & A_{1 2} \\
        A_{2 1} & A_{22}
    \end{pmatrix}
    = \sum_{\text{t} \scriptin \quater} \sigma^{\text{t}} \otimes \cmw_{\text{t} \bm{\ast}}
\end{align}
with \emph{cumulative matrix weights} (CMWs)
\begin{align}\label{equation:CumulativeWeightsDefinition}
\begin{split}
    \cmw_{0 \bm{\ast}} &= \tfrac{1}{2} (A_{11} + A_{22}),\ \cmw_{1 \bm{\ast}} = \tfrac{1}{2} (A_{12} + A_{21}),\\
    \cmw_{2 \bm{\ast}} &= \tfrac{i}{2}(A_{12} - A_{21}),\ \cmw_{3 \bm{\ast}} = \tfrac{1}{2}(A_{11} - A_{22}).
\end{split}
\end{align}
This process is recursively repeated for each of the CMWs, until the remainder strings $\bm{\ast}$ are entirely expanded into concrete strings $\tstring \in \quater^{n}$.
The $n$ levels involved in the full decomposition precisely correspond to applying $(B^{-1})^{\otimes n}$ to the initial matrix $A$.
In particular, notice how the CMWs structure \eqref{equation:CumulativeWeightsDefinition} resembles the rows of $B^{-1}$.
From this new perspective, it is straightforward to define an inverse \texttt{TPD} (\texttt{iTPD}) which transforms a matrix in Pauli representation into the canonical one by applying $B^{\otimes n}$.

Vectorized variants, where the input is a vector of components $A_{j}$ in the canonical basis in standard order are also possible and subsequently used, e.g., for the conversion of the canonical representation to PTM.

\section{\label{section:Methods}Methods}

\subsection{\label{subsection:SpecialSuperoperators}Special superoperators}

We start with some special cases of superoperators that are uniquely determined by one or two $n$-qubit operators in $\Mat(2^{n})$.
In the following, we investigate left/right multiplication with an operator, taking the (anti-)commutator with an operator, and sandwiching with two operators.

\subsubsection{\label{subsubsection:SingleSidedMultiplication}Single-sided multiplication}

Consider an operator $A \in \Mat(2^{n})$ and the two superoperators that are defined by left and right multiplication with $A$, respectively.
We assume $A$ to be available as a matrix w.r.t.\ the canonical basis of $\C^{2^{n}}$.
In order to obtain the \texttt{PTM} of these superoperators we have to study their action on each Pauli string, respectively.
Provided that $A$ is an $n$-fold tensor product of single-qubit operators $A_{i}$, it would be sufficient to record the \texttt{PTM}s of each individual component and to form their Kronecker product (compare tensorality of the \texttt{PTM}).
But even though $A$ will generally not be a tensor product of single-qubit operators, we can apply the same idea that was behind the \texttt{TPD} algorithm \cite{Hantzko2024TensorizedPauliDecompositionAlgorithm}:
recursively writing $A$ as a sum of tensor products of Pauli strings.
Using the linearity of \texttt{PTM}, we can simply add up the results to obtain the \texttt{PTM} of the entire superoperator.
To be more precise, let
\begin{align}\label{equation:CMWDecomposition}
    A = \sum_{\text{t} \scriptin \quater} \sigma^{\text{t}} \otimes \cmw_{\text{t} \bm{\ast}}.
\end{align}
Then multiplying $A$ to a Pauli string $\sigma^{\text{r}} \otimes \sigma^{\sstring}$ from the left yields
\begin{align}\label{equation:LeftMultiplication}
    \Bigg(\sum_{\text{t} \scriptin \quater} \sigma^{\text{t}} \otimes \cmw_{\text{t} \bm{\ast}}\Bigg) (\sigma^{\text{r}} \otimes \sigma^{\sstring}) = \sum_{\text{t} \scriptin \quater} (\sigma^{\text{t}} \sigma^{\text{r}}) \otimes (\cmw_{\text{t} \bm{\ast}} \sigma^{\sstring}).
\end{align}
Analogously, right multiplication with $A$ yields
\begin{align}\label{equation:RightMultiplication}
    \Bigg(\sum_{\text{t} \scriptin \quater} \sigma^{\text{r}} \otimes \sigma^{\sstring}\Bigg) (\sigma^{\text{t}} \otimes \cmw_{\text{t} \bm{\ast}}) = \sum_{\text{t} \scriptin \quater} (\sigma^{\text{r}} \sigma^{\text{t}}) \otimes (\sigma^{\sstring} \cmw_{\text{t} \bm{\ast}}).
\end{align}
The first tensor factors can be pre-computed for all combination of Pauli matrices, yielding the elementary \texttt{PTM}s of left and right multiplication $\epsilon$\texttt{-L/R-PTM} with a Pauli matrix (compare \autoref{table:ElementaryPTMs}).
The arising algorithmic procedure is depicted in \autoref{figure:MultiplicationAlgorithms} (\texttt{L/R-PTM}).

In summary, the \texttt{PTM} of a multiplication with $A$ can be calculated by essentially applying the \texttt{TPD} to $A$ and adding up the tensor products of the contributing elementary \texttt{PTM}s $\epsilon$\texttt{-L/R-PTM}.

\begin{figure*}[!th]
    \begin{minipage}[t]{0.47\linewidth}
        \begin{algorithm}[H]\label{algorithm:CPTM}
            \caption{\texttt{C-PTM}($A$)}
            \eIf{$\dim(A) = 1 \times 1$} {
                \Return{$A$}\;
            } {
                $B \coloneqq 0$\;
                \For{$\text{t} \in \quater$}{
                    compute $\cmw_{\text{t} \bm{\ast}}$ according to \eqref{equation:CumulativeWeightsDefinition}\;
                    \If{$\cmw_{\text{t} \bm{\ast}} \neq 0$} {
                        $B \mathrel{+}= \frac{1}{2} \big(\epsilon\texttt{-C-PTM}[\text{t}] \otimes \texttt{AC-PTM}(\cmw_{\text{t} \bm{\ast}})\big)$\;
                        $B \mathrel{+}= \frac{1}{2} \big(\epsilon\texttt{-AC-PTM}[\text{t}] \otimes \texttt{C-PTM}(\cmw_{\text{t} \bm{\ast}})\big)$\;
                        \tcp{$\epsilon\texttt{-}\text{(}\texttt{A}\text{)}\texttt{C-PTM}[\text{t}]$: \hspace*{-12pt} elementary PTM t}
                        \tcp{for (anti-)commutators}
                    }
                }
                \Return{$B$}\;
            }
        \end{algorithm}
    \end{minipage}\hfill
    \begin{minipage}[t]{0.47\linewidth}
        \begin{algorithm}[H]\label{algorithm:ACPTM}
            \caption{\texttt{AC-PTM}($A$)}
            \eIf{$\dim(A) = 1 \times 1$} {
                \Return{$A$}\;
            } {
                $B \coloneqq 0$\;
                \For{$\text{t} \in \quater$}{
                    compute $\cmw_{\text{t} \bm{\ast}}$ according to \eqref{equation:CumulativeWeightsDefinition}\;
                    \If{$\cmw_{\text{t} \bm{\ast}} \neq 0$} {
                        $B \mathrel{+}= \frac{1}{2} \big(\epsilon\texttt{-C-PTM}[\text{t}] \otimes \texttt{C-PTM}(\cmw_{\text{t} \bm{\ast}})\big)$\;
                        $B \mathrel{+}= \frac{1}{2} \big(\epsilon\texttt{-AC-PTM}[\text{t}] \otimes \texttt{AC-PTM}(\cmw_{\text{t} \bm{\ast}})\big)$\;
                        \tcp{$\epsilon\texttt{-}\text{(}\texttt{A}\text{)}\texttt{C-PTM}[\text{t}]$: \hspace*{-12pt} elementary PTM t}
                        \tcp{for (anti-)commutators}
                    }
                }
                \Return{$B$}\;
            }
        \end{algorithm}
    \end{minipage}
\caption{
    \texttt{PTM}-algorithms for the commutator and anticommutator superoperators.
    Both algorithms input a $2^{n} \times 2^{n}$-matrix $A$ which is simply returned in the case that $n = 0$.
    Otherwise, $B$ is initialized as the zero matrix of dimension $2^{n} \times 2^{n}$ and $A$'s CMWs are calculated via matrix slicing.
    For each nonzero CMW $\cmw_{\text{t} \bm{\ast}}$, two terms added to $B$ which are different for the commutator and the anticommutator case, respectively.
    In the \texttt{C-PTM} algorithm, the halved tensor-product of the t-th element in the lookup table for elementary PTMs for commutators $\epsilon$\texttt{-C-PTM} with the result of \texttt{AC-PTM}($\cmw_{\text{t} \bm{\ast}}$) and of the t-th element in the lookup table for the elementary PTMs for anticommutators $\epsilon$\texttt{-AC-PTM} with the result of \texttt{C-PTM}($\cmw_{\text{t} \bm{\ast}}$) are added to $B$.
    In the \texttt{AC-PTM} algorithm, in contrast, the halved tensor-product of the t-th element in the lookup table for elementary PTMs for commutators $\epsilon$\texttt{-C-PTM} with the result of \texttt{C-PTM}($\cmw_{\text{t} \bm{\ast}}$) and of the t-th element in the lookup table for the elementary PTMs for anticommutators $\epsilon$\texttt{-AC-PTM} with the result of \texttt{AC-PTM}($\cmw_{\text{t} \bm{\ast}}$) are added to $B$.
    Finally, in both algorithms, $B$ is returned.
}
\label{figure:CommutatorAnticommutatorAlgorithms}
\end{figure*}

\subsubsection{\label{subsubsection:CommutatorAndAnticommutator}Commutator and anticommutator}

Consider again an operator $A \in \Mat(2^{n})$ being available as a matrix w.r.t.\ the canonical basis of $\C^{2^{n}}$.
We study next the superoperators defined by forming the commutator and anticommutator with $A$, respectively.
We focus only on the cases where $A$ is the first argument of the (anti-)commutator;
antisymmetry of the commutator and symmetry of the anticommutator readily extend all results to the case where $A$ is the second argument instead.
Assuming again an initial decomposition \eqref{equation:CMWDecomposition} into CMWs, forming the (anti-)commutator of $A$ with a Pauli string $\sigma^{\text{r}} \otimes \sigma^{\sstring}$ yields
\begin{align}
    &\Bigg[\Bigg(\sum_{\text{t} \scriptin \quater} \sigma^{\text{t}} \otimes \cmw_{\text{t} \bm{\ast}}\Bigg), \sigma^{\text{r}} \otimes \sigma^{\sstring}\Bigg] \nonumber \\
    &= \sum_{\text{t} \scriptin \quater} [\sigma^{\text{t}}, \sigma^{\text{r}}] \otimes \cmw_{\text{t} \bm{\ast}} \sigma^{\sstring} + \sigma^{\text{r}} \sigma^{\text{t}} \otimes [\cmw_{\text{t} \bm{\ast}}, \sigma^{\sstring}] \label{equation:CommutatorRelation1} \\
    &= \frac{1}{2}\hspace*{-2pt} \sum_{\text{t} \scriptin \quater} \hspace*{-3pt} \big([\sigma^{\text{t}}, \sigma^{\text{r}}]\hspace*{-0.8pt} \otimes\hspace*{-0.8pt} \{\cmw_{\text{t} \bm{\ast}}, \sigma^{\sstring}\} + \{\sigma^{\text{t}}, \sigma^{\text{r}}\}\hspace*{-0.8pt} \otimes\hspace*{-0.8pt} [\cmw_{\text{t} \bm{\ast}}, \sigma^{\sstring}]\big) \label{equation:CommutatorRelation2}
\end{align}
and
\begin{align}
    &\Bigg\{\Bigg(\sum_{\text{t} \scriptin \quater} \sigma^{\text{t}} \otimes \cmw_{\text{t} \bm{\ast}}\Bigg), \sigma^{\text{r}} \otimes \sigma^{\sstring}\Bigg\} \nonumber \\
    &= \sum_{\text{t} \scriptin \quater} [\sigma^{\text{t}}, \sigma^{\text{r}}] \otimes \cmw_{\text{t} \bm{\ast}} \sigma^{\sstring} + \sigma^{\text{r}} \sigma^{\text{t}} \otimes \{\cmw_{\text{t} \bm{\ast}}, \sigma^{\sstring}\} \label{equation:AnticommutatorRelation1} \\
    &= \frac{1}{2}\hspace*{-2pt} \sum_{\text{t} \scriptin \quater} \hspace*{-3pt} \big([\sigma^{\text{t}}, \sigma^{\text{r}}]\hspace*{-0.8pt} \otimes\hspace*{-0.8pt} [\cmw_{\text{t} \bm{\ast}}, \sigma^{\sstring}] + \{\sigma^{\text{t}}, \sigma^{\text{r}}\}\hspace*{-0.8pt} \otimes\hspace*{-0.8pt} \{\cmw_{\text{t} \bm{\ast}}, \sigma^{\sstring}\}\big) \label{equation:AnticommutatorRelation2}.
\end{align}
The appearing commutator, anticommutators, and products of Pauli matrices can be pre-computed again, forming a set of elementary \texttt{PTM}s $\epsilon$\texttt{-C/AC-PTM} (compare \autoref{table:ElementaryPTMs}).

Similar to the treatment of multiplication superoperators, the calculation of the \texttt{PTM} of forming the commutator or anticommutator with $A$ can be carried out alongside the \texttt{TPD} of $A$ by using the identities \eqref{equation:CommutatorRelation1}--\eqref{equation:AnticommutatorRelation2} and a static lookup table of elementary \texttt{PTM}s.
Pseudo code for both algorithms using \eqref{equation:CommutatorRelation2} and \eqref{equation:AnticommutatorRelation2}, respectively, is presented in \autoref{figure:CommutatorAnticommutatorAlgorithms}.

\subsubsection{\label{subsubsection:SandwichMultiplication}Sandwich multiplication}

\renewcommand*{\arraystretch}{1.5}
\begin{table*}[!th]
    \centering
    \begin{tabularx}{\textwidth}{|>{\centering\arraybackslash}p{1cm}|C|C|C|C|}
    \hline
    \strut & $\sigma^{\text{t}} \defdot $ & $\defdot \sigma^{\text{t}}$ & $[\sigma^{\text{t}}, \defdot]$ & $\{\sigma^{\text{t}}, \defdot\}$ \\
    \hline
    $I$
    & 
    {\renewcommand*{\arraystretch}{1.0}$\begin{pMatrix}{4}
    1 & 0 & 0 & 0 \\
    0 & 1 & 0 & 0 \\
    0 & 0 & 1 & 0 \\
    0 & 0 & 0 & 1
    \end{pMatrix}$}
    & 
    {\renewcommand*{\arraystretch}{1.0}$\begin{pMatrix}{4}
    1 & 0 & 0 & 0 \\
    0 & 1 & 0 & 0 \\
    0 & 0 & 1 & 0 \\
    0 & 0 & 0 & 1
    \end{pMatrix}$}
    &
    {\renewcommand*{\arraystretch}{1.0}$\begin{pMatrix}{4}
    0 & 0 & 0 & 0 \\
    0 & 0 & 0 & 0 \\
    0 & 0 & 0 & 0 \\
    0 & 0 & 0 & 0
    \end{pMatrix}$}
    & 
    {\renewcommand*{\arraystretch}{1.0}$\begin{pMatrix}{4}
    2 & 0 & 0 & 0 \\
    0 & 2 & 0 & 0 \\
    0 & 0 & 2 & 0 \\
    0 & 0 & 0 & 2
    \end{pMatrix}$}
    \\
    \hline
    $X$
    & 
    {\renewcommand*{\arraystretch}{1.0}$\begin{pMatrix}{4}
    0 & 1 & 0 & 0 \\
    1 & 0 & 0 & 0 \\
    0 & 0 & 0 & -i \\
    0 & 0 & i & 0
    \end{pMatrix}$}
    & 
    {\renewcommand*{\arraystretch}{1.0}$\begin{pMatrix}{4}
    0 & 1 & 0 & 0 \\
    1 & 0 & 0 & 0 \\
    0 & 0 & 0 & i \\
    0 & 0 & -i & 0
    \end{pMatrix}$}
    & 
    {\renewcommand*{\arraystretch}{1.0}$\begin{pMatrix}{4}
    0 & 0 & 0 & 0 \\
    0 & 0 & 0 & 0 \\
    0 & 0 & 0 & -2i \\
    0 & 0 & 2i & 0
    \end{pMatrix}$}
    & 
    {\renewcommand*{\arraystretch}{1.0}$\begin{pMatrix}{4}
    0 & 2 & 0 & 0 \\
    2 & 0 & 0 & 0 \\
    0 & 0 & 0 & 0 \\
    0 & 0 & 0 & 0
    \end{pMatrix}$}
    \\
    \hline
    $Y$
    & 
    {\renewcommand*{\arraystretch}{1.0}$\begin{pMatrix}{4}
    0 & 0 & 1 & 0 \\
    0 & 0 & 0 & i \\
    1 & 0 & 0 & 0 \\
    0 & -i & 0 & 0
    \end{pMatrix}$}
    & 
    {\renewcommand*{\arraystretch}{1.0}$\begin{pMatrix}{4}
    0 & 0 & 1 & 0 \\
    0 & 0 & 0 & -i \\
    1 & 0 & 0 & 0 \\
    0 & i & 0 & 0
    \end{pMatrix}$}
    & 
    {\renewcommand*{\arraystretch}{1.0}$\begin{pMatrix}{4}
    0 & 0 & 0 & 0 \\
    0 & 0 & 0 & 2i \\
    0 & 0 & 0 & 0 \\
    0 & -2i & 0 & 0
    \end{pMatrix}$}
    & 
    {\renewcommand*{\arraystretch}{1.0}$\begin{pMatrix}{4}
    0 & 0 & 2 & 0 \\
    0 & 0 & 0 & 0 \\
    2 & 0 & 0 & 0 \\
    0 & 0 & 0 & 0
    \end{pMatrix}$}
    \\
    \hline
    $Z$
    & 
    {\renewcommand*{\arraystretch}{1.0}$\begin{pMatrix}{4}
    0 & 0 & 0 & 1 \\
    0 & 0 & -i & 0 \\
    0 & i & 0 & 0 \\
    1 & 0 & 0 & 0
    \end{pMatrix}$}
    & 
    {\renewcommand*{\arraystretch}{1.0}$\begin{pMatrix}{4}
    0 & 0 & 0 & 1 \\
    0 & 0 & i & 0 \\
    0 & -i & 0 & 0 \\
    1 & 0 & 0 & 0
    \end{pMatrix}$}
    & 
    {\renewcommand*{\arraystretch}{1.0}$\begin{pMatrix}{4}
    0 & 0 & 0 & 0 \\
    0 & 0 & -2i & 0 \\
    0 & 2i & 0 & 0 \\
    0 & 0 & 0 & 0
    \end{pMatrix}$}
    & 
    {\renewcommand*{\arraystretch}{1.0}$\begin{pMatrix}{4}
    0 & 0 & 0 & 2 \\
    0 & 0 & 0 & 0 \\
    0 & 0 & 0 & 0 \\
    2 & 0 & 0 & 0
    \end{pMatrix}$}
    \\
    \hline
    \end{tabularx}
    \caption{Elementary PTMs for the left and right multiplication as well as for the (anti-)commutator superoperators.}
    \label{table:ElementaryPTMs}
\end{table*}
\renewcommand*{\arraystretch}{1.0}

As a final special case we consider the ``sandwich multiplication''-superoperators of the form $[\rho \mapsto A_{1} \rho\, A_{2}]$, where $A_{1}, A_{2} \in \Mat(2^{n})$ are again assumed to be available as matrices w.r.t.\ the canonical basis of $\C^{2^{n}}$.
This class of mappings, in turn, includes pure quantum channels as a special case for $A_{2}^{\vphantom{*}} = A_{1}^{*}$ unitary.
A naive approach to calculate the \texttt{PTM} of such superoperators would be to first calculation the \texttt{PTM} of left multiplication with $A_{1}$, subsequently the \texttt{PTM} of right multiplication with $A_{2}$, and then to multiply both matrices.
However, jointly applying the \texttt{TPD} to both operators $A_{1}$ and $A_{2}$ allows us to calculate the overall \texttt{PTM} more efficiently.
Assuming initial decompositions
\begin{align}\label{equation:CMWDecomposition2}
    A_{i} = \sum_{\text{t} \scriptin \quater} \sigma^{\text{t}} \otimes \cmw_{\text{t} \bm{\ast}}^{(i)},\quad i = 1, 2,
\end{align}
sandwiching a Pauli string $\sigma^{\text{r}} \otimes \sigma^{\sstring}$ between $A_{1}$ and $A_{2}$ yields
\begin{align}\label{equation:SandwichMultiplication}
    &\Bigg(\sum_{\text{t} \scriptin \quater} \sigma^{\text{t}} \otimes \cmw_{\text{t} \bm{\ast}}^{(1)}\Bigg) (\sigma^{\text{r}} \otimes \sigma^{\sstring}) \Bigg(\sum_{\text{u} \scriptin \quater} \sigma^{\text{u}} \otimes \cmw_{\text{u} \bm{\ast}}^{(2)}\Bigg) \nonumber \\
    &\quad = \sum_{\text{t}, \text{u} \in \quater} (\sigma^{\text{t}} \sigma^{\text{r}} \sigma^{\text{u}}) \otimes \Big(\cmw_{\text{t} \bm{\ast}}^{(1)} \sigma^{\sstring} \cmw_{\text{u} \bm{\ast}}^{(2)}\Big).
\end{align}
Again, the first tensor factors can be pre-computed for all $4 \times 4 \times 4 = 64$ possible combinations of Pauli matrices, yielding the elementary \texttt{PTM}s for the sandwich multiplication $\epsilon$\texttt{-M-PTM} (compare \autoref{table:ElementarySandwichMultiplicationPTMs}).
A derived algorithmic primitive is shown in \autoref{figure:MultiplicationAlgorithms} (\texttt{M-PTM}).

\subsection{\label{subsection:ChangeOfChannelRepresentation}Change of channel representation}

Let us now consider the translation from various channel representations such as the (generalized) Kraus representation, the canonical representation, the Choi matrix, and the Chi matrix, to the PTM.

\subsubsection{\label{subsubsection:CanonicalRepresentationToPTM}Canonical representation to PTM}

\renewcommand*{\arraystretch}{1.5}
\begin{table*}[!th]
    \centering
    \begin{tabularx}{\textwidth}{|>{\centering\arraybackslash}p{1cm}|C|C|C|C|}
    \hline
    \strut & $I$ & $X$ & $Y$ & $Z$ \\
    \hline
    $I$
    & 
    {\renewcommand*{\arraystretch}{1.0}$\begin{pMatrix}{4}
    1 & 0 & 0 & 0 \\
    0 & 1 & 0 & 0 \\
    0 & 0 & 1 & 0 \\
    0 & 0 & 0 & 1
    \end{pMatrix}$}
    & 
    {\renewcommand*{\arraystretch}{1.0}$\begin{pMatrix}{4}
    0 & 1 & 0 & 0 \\
    1 & 0 & 0 & 0 \\
    0 & 0 & 0 & i \\
    0 & 0 & -i & 0
    \end{pMatrix}$}
    &
    {\renewcommand*{\arraystretch}{1.0}$\begin{pMatrix}{4}
    0 & 0 & 1 & 0 \\
    0 & 0 & 0 & -i \\
    1 & 0 & 0 & 0 \\
    0 & i & 0 & 0
    \end{pMatrix}$}
    & 
    {\renewcommand*{\arraystretch}{1.0}$\begin{pMatrix}{4}
    0 & 0 & 0 & 1 \\
    0 & 0 & i & 0 \\
    0 & -i & 0 & 0 \\
    1 & 0 & 0 & 0
    \end{pMatrix}$}
    \\
    \hline
    $X$
    & 
    {\renewcommand*{\arraystretch}{1.0}$\begin{pMatrix}{4}
    0 & 1 & 0 & 0 \\
    1 & 0 & 0 & 0 \\
    0 & 0 & 0 & -i \\
    0 & 0 & i & 0
    \end{pMatrix}$}
    & 
    {\renewcommand*{\arraystretch}{1.0}$\begin{pMatrix}{4}
    1 & 0 & 0 & 0 \\
    0 & 1 & 0 & 0 \\
    0 & 0 & -1 & 0 \\
    0 & 0 & 0 & -1
    \end{pMatrix}$}
    & 
    {\renewcommand*{\arraystretch}{1.0}$\begin{pMatrix}{4}
    0 & 0 & 0 & -i \\
    0 & 0 & 1 & 0 \\
    0 & 1 & 0 & 0 \\
    i & 0 & 0 & 0
    \end{pMatrix}$}
    & 
    {\renewcommand*{\arraystretch}{1.0}$\begin{pMatrix}{4}
    0 & 0 & i & 0 \\
    0 & 0 & 0 & 1 \\
    -i & 0 & 0 & 0 \\
    0 & 1 & 0 & 0
    \end{pMatrix}$}
    \\
    \hline
    $Y$
    & 
    {\renewcommand*{\arraystretch}{1.0}$\begin{pMatrix}{4}
    0 & 0 & 1 & 0 \\
    0 & 0 & 0 & i \\
    1 & 0 & 0 & 0 \\
    0 & -i & 0 & 0
    \end{pMatrix}$}
    & 
    {\renewcommand*{\arraystretch}{1.0}$\begin{pMatrix}{4}
    0 & 0 & 0 & i \\
    0 & 0 & 1 & 0 \\
    0 & 1 & 0 & 0 \\
    -i & 0 & 0 & 0
    \end{pMatrix}$}
    & 
    {\renewcommand*{\arraystretch}{1.0}$\begin{pMatrix}{4}
    1 & 0 & 0 & 0 \\
    0 & -1 & 0 & 0 \\
    0 & 0 & 1 & 0 \\
    0 & 0 & 0 & -1
    \end{pMatrix}$}
    & 
    {\renewcommand*{\arraystretch}{1.0}$\begin{pMatrix}{4}
    0 & -i & 0 & 0 \\
    i & 0 & 0 & 0 \\
    0 & 0 & 0 & 1 \\
    0 & 0 & 1 & 0
    \end{pMatrix}$}
    \\
    \hline
    $Z$
    & 
    {\renewcommand*{\arraystretch}{1.0}$\begin{pMatrix}{4}
    0 & 0 & 0 & 1 \\
    0 & 0 & -i & 0 \\
    0 & i & 0 & 0 \\
    1 & 0 & 0 & 0
    \end{pMatrix}$}
    & 
    {\renewcommand*{\arraystretch}{1.0}$\begin{pMatrix}{4}
    0 & 0 & -i & 0 \\
    0 & 0 & 0 & 1 \\
    i & 0 & 0 & 0 \\
    0 & 1 & 0 & 0
    \end{pMatrix}$}
    & 
    {\renewcommand*{\arraystretch}{1.0}$\begin{pMatrix}{4}
    0 & i & 0 & 0 \\
    -i & 0 & 0 & 0 \\
    0 & 0 & 0 & 1 \\
    0 & 0 & 1 & 0
    \end{pMatrix}$}
    & 
    {\renewcommand*{\arraystretch}{1.0}$\begin{pMatrix}{4}
    1 & 0 & 0 & 0 \\
    0 & -1 & 0 & 0 \\
    0 & 0 & -1 & 0 \\
    0 & 0 & 0 & 1
    \end{pMatrix}$}
    \\
    \hline
    \end{tabularx}
    \caption{Elementary PTMs for the sandwich multiplication $[\rho \mapsto \sigma^{\text{t}} \rho\, \sigma^{\text{u}}]$.}
    \label{table:ElementarySandwichMultiplicationPTMs}
\end{table*}
\renewcommand*{\arraystretch}{1.0}

The translation from the canonical representation to PTM corresponds to conducting two basis changes, one along the rows and one along the columns of $\text{Can}(\mathcal{E})$ with its matrix elements $\text{Can}(\mathcal{E})_{i j} =\langle E_{i}, \mathcal{E}(E_{j})\rangle$.
This can be easily seen by recalling that $\text{PTM}(\mathcal{E})_{\sstring \tstring} = \langle\sigma^{\sstring}, \mathcal{E}(\sigma^{\tstring})\rangle$.
Thereby, we have to apply a basis change from the canonical one to the Pauli basis along the rows (realized by applying \texttt{TPD} to the $4^{n}$-dimensional row vectors, reshaped to $2^{n} \times 2^{n}$-matrices) in order to express $\mathcal{E}(E_{j})$ in the Pauli basis, respectively.
Subsequently, we apply the inverse basis change along the columns (realized by applying \texttt{iTPD} to the suitably reshaped column vectors) in order to translate from applying $\mathcal{E}$ to the canonical basis elements to its application to the Pauli strings.
This algorithm is readily invertible by exchanging the application of \texttt{TPD} and \texttt{iTPD}.
Furthermore, since Choi and Chi matrix are related by the very same basis change, the same algorithm(s) may also be applied to translate between these two representations.

\subsubsection{\label{subsubsection:ChoiMatrixToPTM}Choi matrix to PTM}

\begin{figure*}[!th]
    \begin{minipage}[t]{0.47\linewidth}
        \begin{algorithm}[H]\label{algorithm:CanPTM}
            \caption{\texttt{Can-PTM}($M$)}
            $W$ $\coloneqq$ $0$\;
            \For{$i = 1, \ldots, 4^{n}$}{
                $W[i, \,:\,]$ $=$ \texttt{TPD}($M[i, \,:\,]$)\;
                \tcp{vectorized TPD on $i$-th row of $M$}
            }
            $B$ $\coloneqq$ $0$\;
            \For{$j = 1, \ldots, 4^{n}$}{
                $B[\,:\,, j]$ $=$ \texttt{iTPD}($W[\,:\,, j]$)\;
                \tcp{vectorized iTPD on $j$-th column of $W$}
            }
            \Return{$B$}\;
        \end{algorithm}
        \begin{algorithm}[H]\label{algorithm:ChoiPTM}
            \caption{\texttt{Choi-PTM}($M$)}
            $\big\{\cmw_{\tstring \bm{\ast}} \defcolon \tstring \in \quater^{n}\big\}$ $\coloneqq$ $\texttt{iTPD}(M)$\;
            \tcp{interpret $M$ as $2 n$-qubit matrix and}
            \tcp{apply iTPD to first $n$ tensor factors of $M$}
            $B$ $\coloneqq$ $0$\;
            \For{$\tstring \in \quater^{n}$}{
                $B[\,:\,, \tstring] = \texttt{TPD}\big(\cmw_{\tstring \bm{\ast}}\big)$\;
            }
            \Return{$B$}\;
        \end{algorithm}
    \end{minipage}\hfill
    \begin{minipage}[t]{0.47\linewidth}
        \begin{algorithm}[H]\label{algorithm:ChiTM}
            \caption{\texttt{Chi-PTM}($M$)}
            \eIf{$\dim(M) = 1 \times 1$}{
                \Return{$M$}\;
            } {
                $B$ $\coloneqq$ $0$\;
                \For{$\text{s}, \text{t} \in \quater$}{
                    $M_{\text{s}, \text{t}}$ $\coloneqq$ (\text{s},\,\text{t})-block of $M$ \vspace*{1pt}\;
                    \tcp{$M_{\text{s}, \text{t}}$: \hspace*{-12pt} all rows starting with s and}
                    \tcp{all columns starting with t}
                    $B \mathrel{+}= \epsilon\texttt{-M-PTM}[\text{s}][\text{t}] \otimes \texttt{Chi-PTM}(M_{\text{s}, \text{t}})$\;
                    \tcp{$\epsilon\texttt{-M-PTM}[\text{s}][\text{t}]$: \hspace*{-12pt} elementary PTM $(\text{s}, \text{t})$}
                    \tcp{for sandwich multiplication}
                }
            }
            \Return{$B$}\;
        \end{algorithm}
        \begin{algorithm}[H]\label{algorithm:KrausPTM}
            \caption{\texttt{Kraus-PTM}($\{(K_{i}, L_{i})\}_{i}$)}
            $B$ $\coloneqq$ $0$\;
            \For{$(K, L)$ in $\{(K_{i}, L_{i})\}_{i}$ \vspace*{1pt}}{
                $B$ $\mathrel{+}=$ \texttt{M-PTM}($K$, $L^{\dagger}$)\;
            }
            \Return{$B$}\;
        \end{algorithm}
    \end{minipage}
\caption{
    Algorithms for translating from the canonical representation, the Choi matrix, the Chi matrix, and the (generalized) Kraus representation to PTMs, respectively.
    All algorithms except \texttt{Kraus-PTM} input a $4^{n} \times 4^{n}$-matrix;
    \texttt{Kraus-PTM} accepts a list of pairs of $2^{n} \times 2^{n}$-matrices.
    \texttt{Can-PTM} first applies the \texttt{TPD} along the $4^{n}$ rows of the input matrix.
    Subsequently, the \texttt{iTPD} is applied along its $4^{n}$ columns.
    As in both cases (\texttt{i})\texttt{TPD} only operates on a single row/column at once, both steps are highly parallelizable.
    \texttt{Choi-PTM} also employs the \texttt{TPD} and its inverse as subroutines:
    The $4^{n} \times 4^{n}$-dimensional input matrix is interpreted as a $2 n$-qubit matrix and we execute the \texttt{iTPD} on the first $n$ tensor factors, yielding $4^{n}$ partially expanded CMWs.
    Subsequently, we apply the \texttt{TPD} to each of these CMWs and store the results in the columns of the output matrix.
    \texttt{Chi-PTM}, in turn, is defined recursively:
    Return the input if it is a scalar.
    Otherwise, for all eight pairs (s,\,t) of Pauli indices, select the $4^{n - 1} \times 4^{n - 1}$-dimensional sub-matrix $M_{\text{s}, \text{t}}$ whose row indices start with s and column indices start with t and add the Kronecker product of the (s,\,t)-th elementary PTM for sandwich multiplication with \texttt{Chi-PTM}$(M_{\text{s}, \text{t}})$ to the zero-initialized result matrix $B$.
    Lastly, \texttt{Kraus-PTM} iterates over all pairs of (generalized) Kraus operators and adds to result of \texttt{M-PTM} applied to the respective pair to the zero-initialized result $B$. 
}
\label{figure:RepresentationAlgorithms}
\end{figure*}

The inherent tensor structure of the Choi matrix makes it comparatively easy to derive an algorithm for translating it into the PTM.
Recall that by \eqref{equation:ChoiIsomorphism}, the Choi matrix of a quantum channel $\mathcal{E}$ is given by
\begin{equation}
    \text{Choi}(\mathcal{E}) = \sum_{k, \ell = 1}^{2^{n}} E_{k, \ell} \otimes \mathcal{E}(E_{k, \ell}).
\end{equation}
Applying \texttt{iTPD} of depth $n$ to the first tensor factor of $\text{Choi}(\mathcal{E})$ therefore yields the collection of CMWs
\begin{equation}
    \cmw_{\tstring \bm{\ast}} = \sum_{j = 1}^{4^{n}} B_{j \tstring} \mathcal{E}(E_{j}) = \mathcal{E}(\sigma^{\tstring}),\quad \tstring \in \quater^{n}.
\end{equation}
Performing \texttt{TPD} on the individual CMWs then yields the Pauli components of the respective image $\text{PTM}(\mathcal{E})_{\sstring \tstring} = \langle\sigma^{\sstring}, \mathcal{E}(\sigma^{\tstring})\rangle$ with $\sstring \in \quater^{n}$.
These are precisely the $(\sstring, \tstring$)-entries of the PTM which already concludes the algorithmic primitive.

\subsubsection{\label{subsubsection:ChiMatrixToPTM}Chi matrix to PTM}

Converting the Chi matrix into the PTM is similar in spirit to the previously introduced \texttt{M-PTM} algorithm for handling sandwich multiplication.
Recall that the Chi matrix of a quantum channel $\mathcal{E}$ is implicitly defined to satisfy \eqref{equation:ChiMatrix}.
For a given pair $\text{s}, \text{t} \in \quater$, we observe that the (s, t)-blocks of $\text{Chi}(\mathcal{E})$ given by all rows and columns whose Pauli index starts with s and t, respectively, corresponds to the same sandwich multiplication in the first tensor factor: $(\sigma^{\text{t}} \otimes \sigma^{\bm{\ast}}) \rho (\sigma^{\text{s}} \otimes \sigma^{\bm{\ast}})$ and can thus be split off using the elementary sandwich multiplication PTMs $\epsilon$\texttt{-M-PTM} from \autoref{table:ElementarySandwichMultiplicationPTMs}.
Therefore, the PTM for a given $\text{Chi}(\mathcal{E})$ can be obtained recursively by considering all (s,\,t)-blocks and building the Kronecker product of the associated elementary sandwich multiplication PTM with the result of applying the recursive algorithm to the respective (s,\,t)-block.

\subsubsection{\label{subsubsection:GeneralizedKrausRepresentationToPTM}(Generalized) Kraus representation to PTM}

Translating from the (generalized) Kraus representation into PTM is rather straightforward.
Given a set $\{(K_{i}, L_{i})\}_{i} \subset \Mat(2^{n})$ of pairs of generalized Kraus operators (in the ordinary Kraus representation we have that $K_{i} = L_{i}$), we simply add up the PTMs of all individual superoperators $\rho \mapsto K_{i}^{\vphantom{\dagger}} \rho L_{i}^{\dagger}$ with the aid of the previously introduced \texttt{M-PTM} algorithm.

\section{\label{section:Results}Results}

\subsection{\label{subsection:SpecialSuperoperatorsComplexities}Special superoperators: complexities}

All the algorithms \texttt{L/R-PTM}, \texttt{M-PTM}, \texttt{C-PTM}, and \texttt{AC-PTM} can be briefly summarized as applying the \texttt{TPD} algorithm to the input(s) alongside summation of all fully expanded terms where the single-qubit tensor factors can be drawn from a static lookup-table.
Accordingly, the runtime analysis will be quite similar for all these cases and mainly follows the analysis of the worst-case scaling of the \texttt{TPD}~\cite[Section 4]{Hantzko2024TensorizedPauliDecompositionAlgorithm}.

Consider first the \texttt{L/R-PTM} for constructing the PTM of single-sided left/right multiplication with some operator $A \in \Mat(2^{n})$.
Halving the matrix dimensions in every step, the recursion depth of the algorithm is precisely $n$.
At each recursion level $i$, $4^{i}$ CMWs are calculated which each involve taking the sum/difference of two $2^{n- i} \times 2^{n - i}$ matrices, respectively, and the subsequent element-wise division by two.
This accounts for $2 \cdot 4^{i} \cdot 4^{n - i} = 2 \cdot 4^{n}$ operations for the mere calculation of the CMWs per step.
Additionally, for each of the $4^{i}$ Pauli strings of length $i$ at recursion level $i$, a Kronecker product of the respective elementary PTM (a $4 \times 4$-matrix) with the result of the recursively called \texttt{L/R-PTM} has to be calculated and added to the final result $B$.
The matrix returned by \texttt{L/R-PTM} is of dimension $4^{n - i} \times 4^{n - i}$, thus calculating any of the Kronecker products at depth $i$ involves $16 \cdot 16^{n - i} = 16^{n - i + 1}$ many operations.
The same number of operations is incurred when adding the resulting Kronecker product to $B$, introducing a constant factor of $2$.
In total, we obtain 
\begin{align}\label{equation:LeftRightMultiplicationComplexity}
    2 \sum_{i = 1}^{n} 4^{i} \big(4^{n - i} + 16^{n -i + 1}\big) = 2 n 4^{n} + \frac{32}{3} \big(16^{n} - 4^{n}\big)
\end{align}
many elementary operations, i.e.\ $\bigo(16^{n})$ operations.
The $\bigo(n 4^{n})$ operations incurred from calculating the CMWs of $A$ are dominated by the $\bigo(16^{n})$ operations for calculating all the Kronecker products.

Similarly, for both the \texttt{C-PTM} and the \texttt{AC-PTM}, the recursion depth is $n$.
Naively, at each recursion level, the number computational branches increases by a factor of eight (four CMWs times two recursively calculated terms).
However, since $\epsilon$-\texttt{C-PTM}$[0] = 0$ appears in both algorithms, the actual branching factor is only seven.
Additionally, the calculation of the CMWs is the same for both \texttt{C-PTM} and \texttt{AC-PTM} and may therefore be executed externally such that the recursively called \texttt{C-PTM} and \texttt{AC-PTM} do not have to compute them independently.
Finally, for all the $7^{i}$ branches at recursion level $i$, we have to calculate two Kronecker products of an elementary PTM of dimension $4 \times 4$ with a $4^{n - i} \times 4^{n - i}$-matrix, divide the outcomes by two, and add them to the final result $B$.
By the multi-linearity of the Kronecker product, we can simply absorb the factor of $\frac{1}{2}$ into the elementary PTMs which can then also be precomputed.
Analogously to the previous analysis for the \texttt{L/R-PTM}, we obtain a total number of
\begin{align}\label{equation:CommutatorAnticommutatorComplexity}
    2 n 4^{n} + 4 \sum_{i = 1}^{n} 7^{i} \cdot 16^{n - i + 1} = 2 n 4^{n} + \frac{448}{9} \big(16^{n} - 7^{n}\big)
\end{align}
operations, i.e.\ again $\bigo(16^{n})$ operations.
That is, the $\bigo(16^{n})$ operations necessary to calculate the Kronecker products again dominate the $\bigo(n 4^{n})$ operations for calculating the CMWs.

Lastly, we consider the \texttt{M-PTM} for constructing the PTM of sandwich multiplication with two operators $A_{1}, A_{2} \in \Mat(2^{n})$.
Both input matrices' dimensions are halved at each recursion step, yielding again a recursion depth of $n$.
Accounting for all possible pairings of CMWs, the branching factor for the computational tree is $16$ throughout all recursion levels, that is, the $i$-th level possesses $16^{i}$ computational branches.
However, only $2 \cdot 4^{i}$ different CMWs have to be calculated per level (eight in the first step, subsequently four per already computed CMW);
externalizing this task and letting the recursive calls to \texttt{M-PTM} access the results can therefore save many redundant calculations of CMWs.
Similarly to the previously discussed algorithms, the calculation of the Kronecker product at recursion level $i$ of an elementary PTM of dimension $4 \times 4$ with the output of \texttt{M-PTM} of dimension $4^{n - i} \times 4^{n - i}$ and subsequent addition to the final result $B$ necessitate additional $2 \cdot 16^{n - i + 1}$ operations per computational branch.
In summary, this yields
\begin{align}\label{equation:SandwichMultiplicationComplexity}
    4 n 4^{n} + 2 \sum_{i = 1}^{n} 16^{i} 16^{n - i + 1} = 4 n 4^{n} + 32 n 16^{n}
\end{align}
operations.
That is, the \texttt{M-PTM} algorithm admits a runtime of $\bigo(n 16^{n})$, incurring an additional linear factor in comparison to \texttt{L/R-PTM}, \texttt{C-PTM}, and \texttt{AC-PTM}.

\begin{table}[H]
    \centering
    \begin{tabular}{l | l | l}
        algorithm & operations & asymptotic complexity \\\hline
        \texttt{L/R-PTM} & $2 n 4^{n} + \frac{32}{3} (16^{n} - 4^{n})$ & $\bigo(16^{n})$ \\
        \texttt{C/AC-PTM} & $2 n 4^{n} + \frac{448}{9} (16^{n} - 7^{n})$ & $\bigo(16^{n})$ \\
        \texttt{M-PTM} & $4 n 4^{n} + 32 n 16^{n}$ & $\bigo(n 16^{n})$ \\
    \end{tabular}
    \caption{\label{table:SpecialCaseComplexities}Precise operation count and asymptotic complexities for the special case PTM algorithms.}
\end{table}

\subsection{\label{subsection:ChangingChannelRepresentationComplexities}Changing channel representation: complexities}

The number of operations necessary in order to construct the PTM out of a given (generalized) Kraus representation follows directly from the cost of the \texttt{M-PTM} as derived in \eqref{equation:SandwichMultiplicationComplexity}.
For a given collection of $m$ Kraus pairs $\{(K_{i}, L_{i})\}_{i = 1}^{m}$, the total runtime of \texttt{Kraus-PTM} is given by $m$-times the sum of the cost of \texttt{M-PTM} with the cost of adding the respectively obtained $4^{n} \times 4^{n}$ matrix to the final result $B$.
Therefore, \texttt{Kraus-PTM} admits an asymptotic runtime of $\bigo(m n 16^{n})$.

\begin{figure*}[!ht]
    \centering
    \fbox{
    \begin{tabular}{c l}
        \begin{tikzpicture}
            \draw[dashed,QK_plot,line width=2pt] (0,0) to (3,0);
        \end{tikzpicture} & random diagonal matrix \\
        \begin{tikzpicture}
            \draw[QK_plot,line width=2pt] (0,0) to (3,0);
        \end{tikzpicture} & random dense matrix \\
        \begin{tikzpicture}
            \draw[line width=2pt,gray] (0,0) to (3,0);
        \end{tikzpicture} & Slope \\
    \end{tabular}
    } \\
    \input{L-PTM-Tex}
    \input{M-PTM-Tex}
    \input{C-PTM-Tex}
    \input{AC-PTM-Tex}
    \caption{\label{figure:NumericalResultsSpecialCaseAlgorithms}
    Measured execution times of \texttt{L-PTM}, \texttt{M-PTM}, \texttt{C-PTM}, and \texttt{AC-PTM} on random matrices and random diagonal matrices.
    Each plot depicts the respective algorithm's execution time on a logarithmic axis as well as the slope of it's asymptotic runtime (compare \autoref{table:SpecialCaseComplexities}).
    Each algorithm admits an exponential scaling with the number of qubits as derived in the complexity analysis and performs significantly better on random diagonal matrices than on general random matrices.
    The difference in performance is the largest for the \texttt{M-PTM}.}
\end{figure*}

The runtime of \texttt{Can-PTM} immediately follows from the runtimes of \texttt{TPD} and \texttt{iTPD}.
Note that, although we have not given a concrete implementation, the latter may be implemented analogously to \texttt{TPD}, that is, both algorithms require in the worst case $\bigo(n 4^{n})$ many operations.
The \texttt{Can-PTM} algorithm involves two non-nested for-loops of range $4^{n}$ in which \texttt{TPD}/\texttt{iTPD} are applied, respectively.
We conclude an overall runtime of $\bigo(n 16^{n})$.

Similarly, also the runtime of \texttt{Choi-PTM} can easily be derived from the cost of applying the \texttt{TPD} algorithm:
The initial application of \texttt{iTPD} to the first $n$ tensor-factors nevertheless involves handling matrices of dimension $4^{n} \times 4^{n} = 2^{2 n} \times 2^{2 n}$, thus admitting a runtime of $\bigo((2 n) 4^{2 n}) = \bigo(n 16^{n})$.
Subsequently, we have to apply the \texttt{TPD} to $4^{n}$ matrices of dimension $2^{n} \times 2^{n}$, constituting again $\bigo(n 16^{n})$ operations.
Thus, in total, \texttt{Choi-PTM} also requires $\bigo(n 16^{n})$ many operations.

\begin{figure*}
    \centering
    \fbox{
    \begin{tabular}{c l}
        \begin{tikzpicture}
            \draw[dashed,line width=2pt] (0,0) to (3,0);
        \end{tikzpicture} & random diagonal matrix \\
        \begin{tikzpicture}
            \draw[line width=2pt] (0,0) to (3,0);
        \end{tikzpicture} & random matrix \\
        \begin{tikzpicture}
            \draw[line width=2pt,QK_plot] (0,0) to (3,0);
        \end{tikzpicture} & Qiskit \\
        \begin{tikzpicture}
            \draw[line width=2pt,TPD_plot] (0,0) to (3,0);
        \end{tikzpicture} & TPD \\
        \begin{tikzpicture}
            \draw[line width=2pt,HY_plot] (0,0) to (3,0);
        \end{tikzpicture} & Hybrid \\
        \begin{tikzpicture}
            \draw[line width=2pt,gray] (0,0) to (3,0);
        \end{tikzpicture} & Slope \\
    \end{tabular}
    } \\
    \input{Can-PTM-Tex}
    \input{Kraus-PTM-Tex}
    \input{Choi-PTM-Tex}
    \input{Chi-PTM-Tex}
    \caption{\label{figure:NumericalResultsChangingChannelRepresentation}
    Measured execution times of \texttt{Can-PTM}, \texttt{Choi-PTM}, \texttt{Chi}-PTM, the respective conversion algorithms in Qiskit (also for Kraus-to-PTM) as well as hybrid version, utilizing Qiskit to translate to the canonical representation and \texttt{Can-PTM} to translate to the respective PTM.
    The test instances are given by random matrices and random diagonal matrices for systems with up to seven qubits.
    Each plot depicts the respective algorithms' execution time as well as the slope of its asymptotic runtime (compare \autoref{table:ChannelRepresentationComplexities}) on a logarithmic axis.
    In each case, Qiskit's implementation starts out with a favorable runtime.
    On the diagonal test cases, the \texttt{Can-PTM} is able to undercut the former's runtime from six qubits on;
    for general random matrices, \texttt{Can-PTM} requires seven qubits to beat Qiskit' implementation.
    This behavior is reflected in the favorable runtimes of the hybrid method over the respective conversion algorithm in Qiskit for the other three changes of representations.
    For up to five qubits, \texttt{Choi-PTM} and \texttt{Chi-PTM} admit better runtimes than the hybrid versions;
    for six and seven qubits, their runtimes almost match.}
\end{figure*}

Lastly, the \texttt{Chi-PTM}'s runtime can be derived similarly to our analysis for the \texttt{M-PTM}.
At each recursion level, the input matrix' dimensions are divided by four, yielding a recursion depth of $n$ for an input $\text{Chi}(\mathcal{E}) \in \Mat(4^{n})$.
As for the \texttt{M-PTM}, the branching factor for the computational tree is 16 throughout all recursion levels, hence the $i$-th level possesses $16^{i}$ computational branches.
Further analogously, the calculation of the Kronecker product at recursion level $i$ of an elementary PTM of dimension $4 \times 4$ with the output of \texttt{Chi-PTM} of dimension $4^{n - i} \times 4^{n - i}$ and its addition to the final result $B$ incur $\bigo(16^{n - i})$ operations, respectively.
In total, we obtain via analogous reasoning an asymptotic complexity of $\bigo(n 16^{n})$ for the \texttt{Chi-PTM} algorithm.

\begin{table}
    \centering
    \begin{tabular}{l | l}
        algorithm & asymptotic complexity \\\hline
        \texttt{Kraus-PTM} & $\bigo(m n 16^{n})$ \\
        \texttt{Can-PTM} & $\bigo(n 16^{n})$ \\
        \texttt{Choi-PTM} & $\bigo(n 16^{n})$ \\
        \texttt{Chi-PTM} & $\bigo(n 16^{n})$ \\
    \end{tabular}
    \caption{\label{table:ChannelRepresentationComplexities}Comparison of complexities for the generation of Pauli transfer matrices}
\end{table}

\subsection{\label{subsection:SpecialSuperoperatorsNumericalExperiments}Special superoperators: numerical experiments}

In addition to asymptotic complexities, we further present results of numerically testing \texttt{L-PTM}, \texttt{M-PTM}, \texttt{C-PTM}, and \texttt{AC-PTM}.
We are not aware of implementations of algorithms that achieve the same tasks within any (prominent) framework.
Therefore, our numerical experiments for these algorithms merely showcase reasonable practical runtimes.
The results are depicted in \autoref{figure:NumericalResultsSpecialCaseAlgorithms}.
We evaluate the execution times on random matrices and on random diagonal matrices for one to seven qubits.
As expected, the runtimes are overall lower for diagonal matrices which only have up to $2^{n}$ contributing Pauli strings while general random matrices can have up to $4^{n}$ contributing Pauli strings.
The difference between both test sets is most significant for the \texttt{M-PTM}, where the reduced number of contributing Pauli strings enters through both inputs.

\subsection{\label{subsection:ChangingChannelRepresentationNumericalExperiments}Changing channel representation: numerical experiments}

For translating different channel representations into the PTM, we compare our proposed algorithms to implementations found in prominent frameworks such as and Qiskit.
The results are depicted in \autoref{figure:NumericalResultsChangingChannelRepresentation}.
We draw exemplary canonical, Choi and Chi matrices from $4^{n} \times 4^{n}$-dimensional random matrices and random diagonal matrices, and $n$ Kraus operators from $2^{n} \times 2^{n}$-dimensional random matrices and random diagonal matrices.
Those do not necessarily represent a quantum channel, since we do not check the conditions of complete positivity or trace preservation, but are sufficient to be used a benchmarking instances.
For the conversion from Choi/Chi to PTM we test \texttt{Choi-PTM}/\texttt{Chi-PTM} against Qiskit's implementation and a hybrid version, where we translate the Choi/Chi matrix first to the canonical representation via the suitable Qiskit function and then use \texttt{Can-PTM} to translate the canonical representation to the PTM.
When testing \texttt{Can-PTM} itself, the third option is, of course, not applicable.
When generating the data, we observed that the runtime of \texttt{Kraus-PTM} massively exceeded the one of Qiskit's analouge across all qubit numbers.
In order to keep the plot range more compact, we omit the graph of \texttt{Kraus-PTM}'s runtimes and only compare Qiskit's implementation and the hybrid version.
Throughout all test instances we observe that Qiskit's respective conversion algorithm starts out with favorable runtimes, but admits a enormous scaling in the number of qubits.
For random diagonal matrices we find that, for all conversions, either our respective algorithm or the proposed hybrid variant starts to perform better than the pure Qiskit implementation, while admitting a more favorable scaling.
For general random matrices, the required number of qubits for a speed-up of either our algorithm or the hybrid version is seven, again with an improved empirical scaling.

\section{\label{section:ConclusionAndOutlook}Conclusion and Outlook}

Pauli transfer matrices (PTMs) constitute a prominent representation of quantum channels between multi-qubit systems, frequently used in description of quantum processes, quantum tomography etc.
In this article, we have proposed several algorithms for translating other channel representations such as the canonical matrix representation, the Kraus representation, the Choi matrix, and the Chi matrix into PTMs.
Furthermore, we have introduced procedures for calculating the PTMs of special superoperators such as (left, right, and sandwich) multiplication and forming the (anti-)commutator with a multi-qubit operator.
All our methods are essentially based on exploiting the tensor structure of Pauli strings and constitute a rich class of recursive algorithms that manage to detect the input's structure automatically upon execution, resulting in numerically observable reduced runtime.
In addition, our complexity analysis further shows their reasonable worst-case scaling.
From the combination of favorable performance in numerical experiments and the competitive asymptotic scaling, we anticipate improved efficiency for all the tasks covered by our algorithms.

All our algorithms' runtimes are dominated by evaluating the Kronecker product of exponentially large matrices.
However, for at least one of the involved factors, certain symmetry properties are fulfilled which may be utilized to (significantly) speed up the calculation of the respective Kronecker product.
We leave the investigation of such improved methods as an open problem, but wish to highlight their potential, massive impact to our algorithms' runtimes, both in practice and asymptotically.

As a final remark, our algorithmic concepts are mainly centered around the fact that the Pauli basis have a self-replicating tensor structure, that is, e.g., the Pauli basis of a composite multi-qubit system is given by all the tensor products of all the Pauli strings of the component subsystems.
This, in turn, means that all the algorithms may be extended in their scope to also cover other matrix bases with a similar structure such as (generalized) Gell-Mann matrices or Weyl operators.

\begin{acknowledgments}

LB acknowledges financial support by the Quantum Valley Lower Saxony.
LH acknowledges financial support by the Alexander von Humboldt foundation.
We thank Onur Danaci, Tobias J.\ Osborne, Luis Santos, Bence Marton Temesi, and Henrik Wilming for helpful discussions.

\noindent\textbf{Data and code availability statement.}
The depicted data as well as the source code are publicly available under \cite{Repo}.

\end{acknowledgments}

\twocolumngrid

\bibliographystyle{apsrev4-2}
\bibliography{main.bib}

\end{document}

%% file: Trafo.tex
\begin{tikzpicture}
    \begin{scope}[every node/.style={circle,thick,draw}]
        \node (Can) at (-2,0) {Can};
        \node (PTM) at (5,0) {PTM};
        \node (Chi) at (5,5) {Chi};
        \node (Choi) at (-2,5) {Choi};
        \node (Kraus) at (-5,2.5) {Kraus};
    \end{scope}
    \begin{scope}[>={Stealth[black]},
              every node/.style={fill=white,rectangle},
              every edge/.style={draw=JungleGreen,very thick}]
        \path[<->] (Can) edge node {\texttt{(i)Can-PTM}} (PTM);
        \path[<->] (Chi) edge node {\texttt{(i)Can-PTM}} (Choi);
        \path[->] (Choi) edge node {\texttt{Choi-PTM}} (PTM);
        \path[->] (Chi) edge node {\texttt{Chi-PTM}} (PTM);
        \path[dashed, ->] (Kraus) edge node {\texttt{Kraus-PTM}} (PTM);
    \end{scope}
    \begin{scope}[>={Stealth[black]},
              every node/.style={fill=white,rectangle},
              every edge/.style={draw=black,very thick}]
        \path[<->] (Kraus) edge node {Qiskit} (Choi);
        \path[->] (Kraus) edge node {Qiskit} (Can);
        \path[<->] (Can) edge node {Qiskit} (Choi);
    \end{scope}
\end{tikzpicture}

%% file: L-PTM-Tex.tex
\begingroup
  \makeatletter
  \providecommand\color[2][]{%
    \GenericError{(gnuplot) \space\space\space\@spaces}{%
      Package color not loaded in conjunction with
      terminal option `colourtext'%
    }{See the gnuplot documentation for explanation.%
    }{Either use 'blacktext' in gnuplot or load the package
      color.sty in LaTeX.}%
    \renewcommand\color[2][]{}%
  }%
  \providecommand\includegraphics[2][]{%
    \GenericError{(gnuplot) \space\space\space\@spaces}{%
      Package graphicx or graphics not loaded%
    }{See the gnuplot documentation for explanation.%
    }{The gnuplot epslatex terminal needs graphicx.sty or graphics.sty.}%
    \renewcommand\includegraphics[2][]{}%
  }%
  \providecommand\rotatebox[2]{#2}%
  \@ifundefined{ifGPcolor}{%
    \newif\ifGPcolor
    \GPcolortrue
  }{}%
  \@ifundefined{ifGPblacktext}{%
    \newif\ifGPblacktext
    \GPblacktextfalse
  }{}%
  \let\gplgaddtomacro\g@addto@macro
  \gdef\gplbacktext{}%
  \gdef\gplfronttext{}%
  \makeatother
  \ifGPblacktext
    \def\colorrgb#1{}%
    \def\colorgray#1{}%
  \else
    \ifGPcolor
      \def\colorrgb#1{\color[rgb]{#1}}%
      \def\colorgray#1{\color[gray]{#1}}%
      \expandafter\def\csname LTw\endcsname{\color{white}}%
      \expandafter\def\csname LTb\endcsname{\color{black}}%
      \expandafter\def\csname LTa\endcsname{\color{black}}%
      \expandafter\def\csname LT0\endcsname{\color[rgb]{1,0,0}}%
      \expandafter\def\csname LT1\endcsname{\color[rgb]{0,1,0}}%
      \expandafter\def\csname LT2\endcsname{\color[rgb]{0,0,1}}%
      \expandafter\def\csname LT3\endcsname{\color[rgb]{1,0,1}}%
      \expandafter\def\csname LT4\endcsname{\color[rgb]{0,1,1}}%
      \expandafter\def\csname LT5\endcsname{\color[rgb]{1,1,0}}%
      \expandafter\def\csname LT6\endcsname{\color[rgb]{0,0,0}}%
      \expandafter\def\csname LT7\endcsname{\color[rgb]{1,0.3,0}}%
      \expandafter\def\csname LT8\endcsname{\color[rgb]{0.5,0.5,0.5}}%
    \else
      \def\colorrgb#1{\color{black}}%
      \def\colorgray#1{\color[gray]{#1}}%
      \expandafter\def\csname LTw\endcsname{\color{white}}%
      \expandafter\def\csname LTb\endcsname{\color{black}}%
      \expandafter\def\csname LTa\endcsname{\color{black}}%
      \expandafter\def\csname LT0\endcsname{\color{black}}%
      \expandafter\def\csname LT1\endcsname{\color{black}}%
      \expandafter\def\csname LT2\endcsname{\color{black}}%
      \expandafter\def\csname LT3\endcsname{\color{black}}%
      \expandafter\def\csname LT4\endcsname{\color{black}}%
      \expandafter\def\csname LT5\endcsname{\color{black}}%
      \expandafter\def\csname LT6\endcsname{\color{black}}%
      \expandafter\def\csname LT7\endcsname{\color{black}}%
      \expandafter\def\csname LT8\endcsname{\color{black}}%
    \fi
  \fi
    \setlength{\unitlength}{0.0500bp}%
    \ifx\gptboxheight\undefined%
      \newlength{\gptboxheight}%
      \newlength{\gptboxwidth}%
      \newsavebox{\gptboxtext}%
    \fi%
    \setlength{\fboxrule}{0.5pt}%
    \setlength{\fboxsep}{1pt}%
    \definecolor{tbcol}{rgb}{1,1,1}%
\begin{picture}(4520.00,4520.00)%
    \gplgaddtomacro\gplbacktext{%
      \colorrgb{0.50,0.50,0.50}
      \put(594,834){\makebox(0,0)[r]{\strut{}$10^{-6}$}}%
      \colorrgb{0.50,0.50,0.50}
      \put(594,1549){\makebox(0,0)[r]{\strut{}$10^{-4}$}}%
      \colorrgb{0.50,0.50,0.50}
      \put(594,2264){\makebox(0,0)[r]{\strut{}$10^{-2}$}}%
      \colorrgb{0.50,0.50,0.50}
      \put(594,2979){\makebox(0,0)[r]{\strut{}$10^{0}$}}%
      \colorrgb{0.50,0.50,0.50}
      \put(594,3694){\makebox(0,0)[r]{\strut{}$10^{2}$}}%
      \colorrgb{0.50,0.50,0.50}
      \put(730,327){\makebox(0,0){\strut{}$1$}}%
      \colorrgb{0.50,0.50,0.50}
      \put(1282,327){\makebox(0,0){\strut{}$2$}}%
      \colorrgb{0.50,0.50,0.50}
      \put(1833,327){\makebox(0,0){\strut{}$3$}}%
      \colorrgb{0.50,0.50,0.50}
      \put(2385,327){\makebox(0,0){\strut{}$4$}}%
      \colorrgb{0.50,0.50,0.50}
      \put(2936,327){\makebox(0,0){\strut{}$5$}}%
      \colorrgb{0.50,0.50,0.50}
      \put(3487,327){\makebox(0,0){\strut{}$6$}}%
      \colorrgb{0.50,0.50,0.50}
      \put(4039,327){\makebox(0,0){\strut{}$7$}}%
    }%
    \gplgaddtomacro\gplfronttext{%
      \csname LTb\endcsname
      \put(135,2264){\rotatebox{-270}{\makebox(0,0){\strut{}Execution Time $s$}}}%
      \csname LTb\endcsname
      \put(2384,104){\makebox(0,0){\strut{}Number of Qubits $n$}}%
      \csname LTb\endcsname
      \put(2384,4275){\makebox(0,0){\strut{}\texttt{L-PTM}}}%
    }%
    \gplbacktext
    \put(0,0){\includegraphics[width={226.00bp},height={226.00bp}]{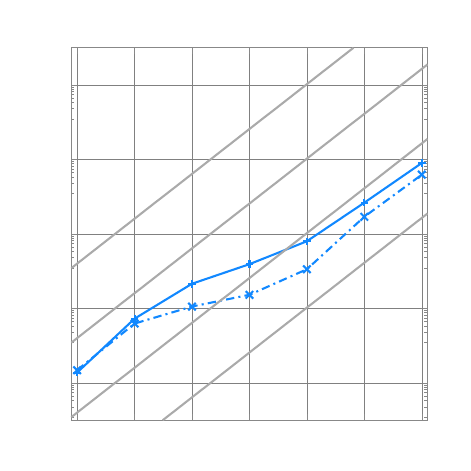}}%
    \gplfronttext
  \end{picture}%
\endgroup

%% file: M-PTM-Tex.tex
\begingroup
  \makeatletter
  \providecommand\color[2][]{%
    \GenericError{(gnuplot) \space\space\space\@spaces}{%
      Package color not loaded in conjunction with
      terminal option `colourtext'%
    }{See the gnuplot documentation for explanation.%
    }{Either use 'blacktext' in gnuplot or load the package
      color.sty in LaTeX.}%
    \renewcommand\color[2][]{}%
  }%
  \providecommand\includegraphics[2][]{%
    \GenericError{(gnuplot) \space\space\space\@spaces}{%
      Package graphicx or graphics not loaded%
    }{See the gnuplot documentation for explanation.%
    }{The gnuplot epslatex terminal needs graphicx.sty or graphics.sty.}%
    \renewcommand\includegraphics[2][]{}%
  }%
  \providecommand\rotatebox[2]{#2}%
  \@ifundefined{ifGPcolor}{%
    \newif\ifGPcolor
    \GPcolortrue
  }{}%
  \@ifundefined{ifGPblacktext}{%
    \newif\ifGPblacktext
    \GPblacktextfalse
  }{}%
  \let\gplgaddtomacro\g@addto@macro
  \gdef\gplbacktext{}%
  \gdef\gplfronttext{}%
  \makeatother
  \ifGPblacktext
    \def\colorrgb#1{}%
    \def\colorgray#1{}%
  \else
    \ifGPcolor
      \def\colorrgb#1{\color[rgb]{#1}}%
      \def\colorgray#1{\color[gray]{#1}}%
      \expandafter\def\csname LTw\endcsname{\color{white}}%
      \expandafter\def\csname LTb\endcsname{\color{black}}%
      \expandafter\def\csname LTa\endcsname{\color{black}}%
      \expandafter\def\csname LT0\endcsname{\color[rgb]{1,0,0}}%
      \expandafter\def\csname LT1\endcsname{\color[rgb]{0,1,0}}%
      \expandafter\def\csname LT2\endcsname{\color[rgb]{0,0,1}}%
      \expandafter\def\csname LT3\endcsname{\color[rgb]{1,0,1}}%
      \expandafter\def\csname LT4\endcsname{\color[rgb]{0,1,1}}%
      \expandafter\def\csname LT5\endcsname{\color[rgb]{1,1,0}}%
      \expandafter\def\csname LT6\endcsname{\color[rgb]{0,0,0}}%
      \expandafter\def\csname LT7\endcsname{\color[rgb]{1,0.3,0}}%
      \expandafter\def\csname LT8\endcsname{\color[rgb]{0.5,0.5,0.5}}%
    \else
      \def\colorrgb#1{\color{black}}%
      \def\colorgray#1{\color[gray]{#1}}%
      \expandafter\def\csname LTw\endcsname{\color{white}}%
      \expandafter\def\csname LTb\endcsname{\color{black}}%
      \expandafter\def\csname LTa\endcsname{\color{black}}%
      \expandafter\def\csname LT0\endcsname{\color{black}}%
      \expandafter\def\csname LT1\endcsname{\color{black}}%
      \expandafter\def\csname LT2\endcsname{\color{black}}%
      \expandafter\def\csname LT3\endcsname{\color{black}}%
      \expandafter\def\csname LT4\endcsname{\color{black}}%
      \expandafter\def\csname LT5\endcsname{\color{black}}%
      \expandafter\def\csname LT6\endcsname{\color{black}}%
      \expandafter\def\csname LT7\endcsname{\color{black}}%
      \expandafter\def\csname LT8\endcsname{\color{black}}%
    \fi
  \fi
    \setlength{\unitlength}{0.0500bp}%
    \ifx\gptboxheight\undefined%
      \newlength{\gptboxheight}%
      \newlength{\gptboxwidth}%
      \newsavebox{\gptboxtext}%
    \fi%
    \setlength{\fboxrule}{0.5pt}%
    \setlength{\fboxsep}{1pt}%
    \definecolor{tbcol}{rgb}{1,1,1}%
\begin{picture}(4520.00,4520.00)%
    \gplgaddtomacro\gplbacktext{%
      \colorrgb{0.50,0.50,0.50}
      \put(594,834){\makebox(0,0)[r]{\strut{}$10^{-6}$}}%
      \colorrgb{0.50,0.50,0.50}
      \put(594,1549){\makebox(0,0)[r]{\strut{}$10^{-4}$}}%
      \colorrgb{0.50,0.50,0.50}
      \put(594,2264){\makebox(0,0)[r]{\strut{}$10^{-2}$}}%
      \colorrgb{0.50,0.50,0.50}
      \put(594,2979){\makebox(0,0)[r]{\strut{}$10^{0}$}}%
      \colorrgb{0.50,0.50,0.50}
      \put(594,3694){\makebox(0,0)[r]{\strut{}$10^{2}$}}%
      \colorrgb{0.50,0.50,0.50}
      \put(730,327){\makebox(0,0){\strut{}$1$}}%
      \colorrgb{0.50,0.50,0.50}
      \put(1282,327){\makebox(0,0){\strut{}$2$}}%
      \colorrgb{0.50,0.50,0.50}
      \put(1833,327){\makebox(0,0){\strut{}$3$}}%
      \colorrgb{0.50,0.50,0.50}
      \put(2385,327){\makebox(0,0){\strut{}$4$}}%
      \colorrgb{0.50,0.50,0.50}
      \put(2936,327){\makebox(0,0){\strut{}$5$}}%
      \colorrgb{0.50,0.50,0.50}
      \put(3487,327){\makebox(0,0){\strut{}$6$}}%
      \colorrgb{0.50,0.50,0.50}
      \put(4039,327){\makebox(0,0){\strut{}$7$}}%
    }%
    \gplgaddtomacro\gplfronttext{%
      \csname LTb\endcsname
      \put(135,2264){\rotatebox{-270}{\makebox(0,0){\strut{}Execution Time $s$}}}%
      \csname LTb\endcsname
      \put(2384,104){\makebox(0,0){\strut{}Number of Qubits $n$}}%
      \csname LTb\endcsname
      \put(2384,4275){\makebox(0,0){\strut{}\texttt{M-PTM}}}%
    }%
    \gplbacktext
    \put(0,0){\includegraphics[width={226.00bp},height={226.00bp}]{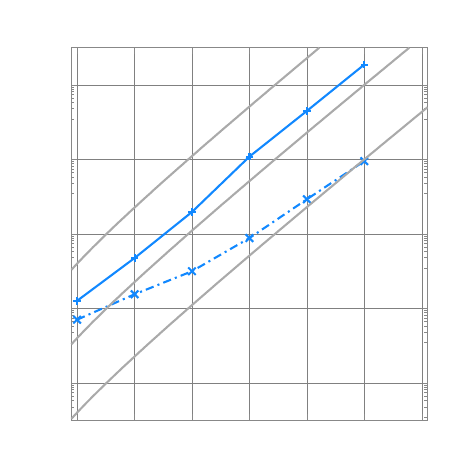}}%
    \gplfronttext
  \end{picture}%
\endgroup

%% file: C-PTM-Tex.tex
\begingroup
  \makeatletter
  \providecommand\color[2][]{%
    \GenericError{(gnuplot) \space\space\space\@spaces}{%
      Package color not loaded in conjunction with
      terminal option `colourtext'%
    }{See the gnuplot documentation for explanation.%
    }{Either use 'blacktext' in gnuplot or load the package
      color.sty in LaTeX.}%
    \renewcommand\color[2][]{}%
  }%
  \providecommand\includegraphics[2][]{%
    \GenericError{(gnuplot) \space\space\space\@spaces}{%
      Package graphicx or graphics not loaded%
    }{See the gnuplot documentation for explanation.%
    }{The gnuplot epslatex terminal needs graphicx.sty or graphics.sty.}%
    \renewcommand\includegraphics[2][]{}%
  }%
  \providecommand\rotatebox[2]{#2}%
  \@ifundefined{ifGPcolor}{%
    \newif\ifGPcolor
    \GPcolortrue
  }{}%
  \@ifundefined{ifGPblacktext}{%
    \newif\ifGPblacktext
    \GPblacktextfalse
  }{}%
  \let\gplgaddtomacro\g@addto@macro
  \gdef\gplbacktext{}%
  \gdef\gplfronttext{}%
  \makeatother
  \ifGPblacktext
    \def\colorrgb#1{}%
    \def\colorgray#1{}%
  \else
    \ifGPcolor
      \def\colorrgb#1{\color[rgb]{#1}}%
      \def\colorgray#1{\color[gray]{#1}}%
      \expandafter\def\csname LTw\endcsname{\color{white}}%
      \expandafter\def\csname LTb\endcsname{\color{black}}%
      \expandafter\def\csname LTa\endcsname{\color{black}}%
      \expandafter\def\csname LT0\endcsname{\color[rgb]{1,0,0}}%
      \expandafter\def\csname LT1\endcsname{\color[rgb]{0,1,0}}%
      \expandafter\def\csname LT2\endcsname{\color[rgb]{0,0,1}}%
      \expandafter\def\csname LT3\endcsname{\color[rgb]{1,0,1}}%
      \expandafter\def\csname LT4\endcsname{\color[rgb]{0,1,1}}%
      \expandafter\def\csname LT5\endcsname{\color[rgb]{1,1,0}}%
      \expandafter\def\csname LT6\endcsname{\color[rgb]{0,0,0}}%
      \expandafter\def\csname LT7\endcsname{\color[rgb]{1,0.3,0}}%
      \expandafter\def\csname LT8\endcsname{\color[rgb]{0.5,0.5,0.5}}%
    \else
      \def\colorrgb#1{\color{black}}%
      \def\colorgray#1{\color[gray]{#1}}%
      \expandafter\def\csname LTw\endcsname{\color{white}}%
      \expandafter\def\csname LTb\endcsname{\color{black}}%
      \expandafter\def\csname LTa\endcsname{\color{black}}%
      \expandafter\def\csname LT0\endcsname{\color{black}}%
      \expandafter\def\csname LT1\endcsname{\color{black}}%
      \expandafter\def\csname LT2\endcsname{\color{black}}%
      \expandafter\def\csname LT3\endcsname{\color{black}}%
      \expandafter\def\csname LT4\endcsname{\color{black}}%
      \expandafter\def\csname LT5\endcsname{\color{black}}%
      \expandafter\def\csname LT6\endcsname{\color{black}}%
      \expandafter\def\csname LT7\endcsname{\color{black}}%
      \expandafter\def\csname LT8\endcsname{\color{black}}%
    \fi
  \fi
    \setlength{\unitlength}{0.0500bp}%
    \ifx\gptboxheight\undefined%
      \newlength{\gptboxheight}%
      \newlength{\gptboxwidth}%
      \newsavebox{\gptboxtext}%
    \fi%
    \setlength{\fboxrule}{0.5pt}%
    \setlength{\fboxsep}{1pt}%
    \definecolor{tbcol}{rgb}{1,1,1}%
\begin{picture}(4520.00,4520.00)%
    \gplgaddtomacro\gplbacktext{%
      \colorrgb{0.50,0.50,0.50}
      \put(594,834){\makebox(0,0)[r]{\strut{}$10^{-6}$}}%
      \colorrgb{0.50,0.50,0.50}
      \put(594,1549){\makebox(0,0)[r]{\strut{}$10^{-4}$}}%
      \colorrgb{0.50,0.50,0.50}
      \put(594,2264){\makebox(0,0)[r]{\strut{}$10^{-2}$}}%
      \colorrgb{0.50,0.50,0.50}
      \put(594,2979){\makebox(0,0)[r]{\strut{}$10^{0}$}}%
      \colorrgb{0.50,0.50,0.50}
      \put(594,3694){\makebox(0,0)[r]{\strut{}$10^{2}$}}%
      \colorrgb{0.50,0.50,0.50}
      \put(730,327){\makebox(0,0){\strut{}$1$}}%
      \colorrgb{0.50,0.50,0.50}
      \put(1282,327){\makebox(0,0){\strut{}$2$}}%
      \colorrgb{0.50,0.50,0.50}
      \put(1833,327){\makebox(0,0){\strut{}$3$}}%
      \colorrgb{0.50,0.50,0.50}
      \put(2385,327){\makebox(0,0){\strut{}$4$}}%
      \colorrgb{0.50,0.50,0.50}
      \put(2936,327){\makebox(0,0){\strut{}$5$}}%
      \colorrgb{0.50,0.50,0.50}
      \put(3487,327){\makebox(0,0){\strut{}$6$}}%
      \colorrgb{0.50,0.50,0.50}
      \put(4039,327){\makebox(0,0){\strut{}$7$}}%
    }%
    \gplgaddtomacro\gplfronttext{%
      \csname LTb\endcsname
      \put(135,2264){\rotatebox{-270}{\makebox(0,0){\strut{}Execution Time $s$}}}%
      \csname LTb\endcsname
      \put(2384,104){\makebox(0,0){\strut{}Number of Qubits $n$}}%
      \csname LTb\endcsname
      \put(2384,4275){\makebox(0,0){\strut{}\texttt{C-PTM}}}%
    }%
    \gplbacktext
    \put(0,0){\includegraphics[width={226.00bp},height={226.00bp}]{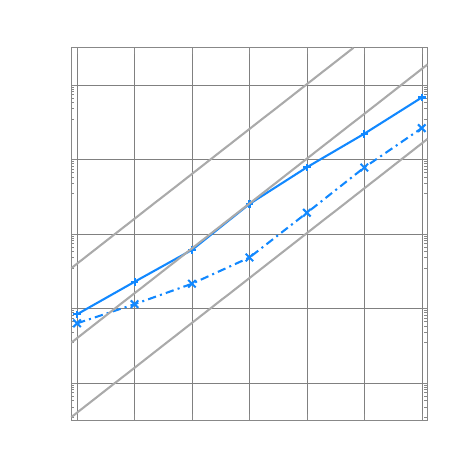}}%
    \gplfronttext
  \end{picture}%
\endgroup

%% file: AC-PTM-Tex.tex
\begingroup
  \makeatletter
  \providecommand\color[2][]{%
    \GenericError{(gnuplot) \space\space\space\@spaces}{%
      Package color not loaded in conjunction with
      terminal option `colourtext'%
    }{See the gnuplot documentation for explanation.%
    }{Either use 'blacktext' in gnuplot or load the package
      color.sty in LaTeX.}%
    \renewcommand\color[2][]{}%
  }%
  \providecommand\includegraphics[2][]{%
    \GenericError{(gnuplot) \space\space\space\@spaces}{%
      Package graphicx or graphics not loaded%
    }{See the gnuplot documentation for explanation.%
    }{The gnuplot epslatex terminal needs graphicx.sty or graphics.sty.}%
    \renewcommand\includegraphics[2][]{}%
  }%
  \providecommand\rotatebox[2]{#2}%
  \@ifundefined{ifGPcolor}{%
    \newif\ifGPcolor
    \GPcolortrue
  }{}%
  \@ifundefined{ifGPblacktext}{%
    \newif\ifGPblacktext
    \GPblacktextfalse
  }{}%
  \let\gplgaddtomacro\g@addto@macro
  \gdef\gplbacktext{}%
  \gdef\gplfronttext{}%
  \makeatother
  \ifGPblacktext
    \def\colorrgb#1{}%
    \def\colorgray#1{}%
  \else
    \ifGPcolor
      \def\colorrgb#1{\color[rgb]{#1}}%
      \def\colorgray#1{\color[gray]{#1}}%
      \expandafter\def\csname LTw\endcsname{\color{white}}%
      \expandafter\def\csname LTb\endcsname{\color{black}}%
      \expandafter\def\csname LTa\endcsname{\color{black}}%
      \expandafter\def\csname LT0\endcsname{\color[rgb]{1,0,0}}%
      \expandafter\def\csname LT1\endcsname{\color[rgb]{0,1,0}}%
      \expandafter\def\csname LT2\endcsname{\color[rgb]{0,0,1}}%
      \expandafter\def\csname LT3\endcsname{\color[rgb]{1,0,1}}%
      \expandafter\def\csname LT4\endcsname{\color[rgb]{0,1,1}}%
      \expandafter\def\csname LT5\endcsname{\color[rgb]{1,1,0}}%
      \expandafter\def\csname LT6\endcsname{\color[rgb]{0,0,0}}%
      \expandafter\def\csname LT7\endcsname{\color[rgb]{1,0.3,0}}%
      \expandafter\def\csname LT8\endcsname{\color[rgb]{0.5,0.5,0.5}}%
    \else
      \def\colorrgb#1{\color{black}}%
      \def\colorgray#1{\color[gray]{#1}}%
      \expandafter\def\csname LTw\endcsname{\color{white}}%
      \expandafter\def\csname LTb\endcsname{\color{black}}%
      \expandafter\def\csname LTa\endcsname{\color{black}}%
      \expandafter\def\csname LT0\endcsname{\color{black}}%
      \expandafter\def\csname LT1\endcsname{\color{black}}%
      \expandafter\def\csname LT2\endcsname{\color{black}}%
      \expandafter\def\csname LT3\endcsname{\color{black}}%
      \expandafter\def\csname LT4\endcsname{\color{black}}%
      \expandafter\def\csname LT5\endcsname{\color{black}}%
      \expandafter\def\csname LT6\endcsname{\color{black}}%
      \expandafter\def\csname LT7\endcsname{\color{black}}%
      \expandafter\def\csname LT8\endcsname{\color{black}}%
    \fi
  \fi
    \setlength{\unitlength}{0.0500bp}%
    \ifx\gptboxheight\undefined%
      \newlength{\gptboxheight}%
      \newlength{\gptboxwidth}%
      \newsavebox{\gptboxtext}%
    \fi%
    \setlength{\fboxrule}{0.5pt}%
    \setlength{\fboxsep}{1pt}%
    \definecolor{tbcol}{rgb}{1,1,1}%
\begin{picture}(4520.00,4520.00)%
    \gplgaddtomacro\gplbacktext{%
      \colorrgb{0.50,0.50,0.50}
      \put(594,834){\makebox(0,0)[r]{\strut{}$10^{-6}$}}%
      \colorrgb{0.50,0.50,0.50}
      \put(594,1549){\makebox(0,0)[r]{\strut{}$10^{-4}$}}%
      \colorrgb{0.50,0.50,0.50}
      \put(594,2264){\makebox(0,0)[r]{\strut{}$10^{-2}$}}%
      \colorrgb{0.50,0.50,0.50}
      \put(594,2979){\makebox(0,0)[r]{\strut{}$10^{0}$}}%
      \colorrgb{0.50,0.50,0.50}
      \put(594,3694){\makebox(0,0)[r]{\strut{}$10^{2}$}}%
      \colorrgb{0.50,0.50,0.50}
      \put(730,327){\makebox(0,0){\strut{}$1$}}%
      \colorrgb{0.50,0.50,0.50}
      \put(1282,327){\makebox(0,0){\strut{}$2$}}%
      \colorrgb{0.50,0.50,0.50}
      \put(1833,327){\makebox(0,0){\strut{}$3$}}%
      \colorrgb{0.50,0.50,0.50}
      \put(2385,327){\makebox(0,0){\strut{}$4$}}%
      \colorrgb{0.50,0.50,0.50}
      \put(2936,327){\makebox(0,0){\strut{}$5$}}%
      \colorrgb{0.50,0.50,0.50}
      \put(3487,327){\makebox(0,0){\strut{}$6$}}%
      \colorrgb{0.50,0.50,0.50}
      \put(4039,327){\makebox(0,0){\strut{}$7$}}%
    }%
    \gplgaddtomacro\gplfronttext{%
      \csname LTb\endcsname
      \put(135,2264){\rotatebox{-270}{\makebox(0,0){\strut{}Execution Time $s$}}}%
      \csname LTb\endcsname
      \put(2384,104){\makebox(0,0){\strut{}Number of Qubits $n$}}%
      \csname LTb\endcsname
      \put(2384,4275){\makebox(0,0){\strut{}\texttt{AC-PTM}}}%
    }%
    \gplbacktext
    \put(0,0){\includegraphics[width={226.00bp},height={226.00bp}]{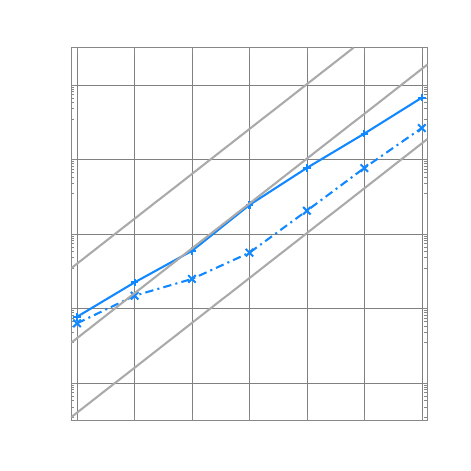}}%
    \gplfronttext
  \end{picture}%
\endgroup

%% file: Can-PTM-Tex.tex
\begingroup
  \makeatletter
  \providecommand\color[2][]{%
    \GenericError{(gnuplot) \space\space\space\@spaces}{%
      Package color not loaded in conjunction with
      terminal option `colourtext'%
    }{See the gnuplot documentation for explanation.%
    }{Either use 'blacktext' in gnuplot or load the package
      color.sty in LaTeX.}%
    \renewcommand\color[2][]{}%
  }%
  \providecommand\includegraphics[2][]{%
    \GenericError{(gnuplot) \space\space\space\@spaces}{%
      Package graphicx or graphics not loaded%
    }{See the gnuplot documentation for explanation.%
    }{The gnuplot epslatex terminal needs graphicx.sty or graphics.sty.}%
    \renewcommand\includegraphics[2][]{}%
  }%
  \providecommand\rotatebox[2]{#2}%
  \@ifundefined{ifGPcolor}{%
    \newif\ifGPcolor
    \GPcolortrue
  }{}%
  \@ifundefined{ifGPblacktext}{%
    \newif\ifGPblacktext
    \GPblacktextfalse
  }{}%
  \let\gplgaddtomacro\g@addto@macro
  \gdef\gplbacktext{}%
  \gdef\gplfronttext{}%
  \makeatother
  \ifGPblacktext
    \def\colorrgb#1{}%
    \def\colorgray#1{}%
  \else
    \ifGPcolor
      \def\colorrgb#1{\color[rgb]{#1}}%
      \def\colorgray#1{\color[gray]{#1}}%
      \expandafter\def\csname LTw\endcsname{\color{white}}%
      \expandafter\def\csname LTb\endcsname{\color{black}}%
      \expandafter\def\csname LTa\endcsname{\color{black}}%
      \expandafter\def\csname LT0\endcsname{\color[rgb]{1,0,0}}%
      \expandafter\def\csname LT1\endcsname{\color[rgb]{0,1,0}}%
      \expandafter\def\csname LT2\endcsname{\color[rgb]{0,0,1}}%
      \expandafter\def\csname LT3\endcsname{\color[rgb]{1,0,1}}%
      \expandafter\def\csname LT4\endcsname{\color[rgb]{0,1,1}}%
      \expandafter\def\csname LT5\endcsname{\color[rgb]{1,1,0}}%
      \expandafter\def\csname LT6\endcsname{\color[rgb]{0,0,0}}%
      \expandafter\def\csname LT7\endcsname{\color[rgb]{1,0.3,0}}%
      \expandafter\def\csname LT8\endcsname{\color[rgb]{0.5,0.5,0.5}}%
    \else
      \def\colorrgb#1{\color{black}}%
      \def\colorgray#1{\color[gray]{#1}}%
      \expandafter\def\csname LTw\endcsname{\color{white}}%
      \expandafter\def\csname LTb\endcsname{\color{black}}%
      \expandafter\def\csname LTa\endcsname{\color{black}}%
      \expandafter\def\csname LT0\endcsname{\color{black}}%
      \expandafter\def\csname LT1\endcsname{\color{black}}%
      \expandafter\def\csname LT2\endcsname{\color{black}}%
      \expandafter\def\csname LT3\endcsname{\color{black}}%
      \expandafter\def\csname LT4\endcsname{\color{black}}%
      \expandafter\def\csname LT5\endcsname{\color{black}}%
      \expandafter\def\csname LT6\endcsname{\color{black}}%
      \expandafter\def\csname LT7\endcsname{\color{black}}%
      \expandafter\def\csname LT8\endcsname{\color{black}}%
    \fi
  \fi
    \setlength{\unitlength}{0.0500bp}%
    \ifx\gptboxheight\undefined%
      \newlength{\gptboxheight}%
      \newlength{\gptboxwidth}%
      \newsavebox{\gptboxtext}%
    \fi%
    \setlength{\fboxrule}{0.5pt}%
    \setlength{\fboxsep}{1pt}%
    \definecolor{tbcol}{rgb}{1,1,1}%
\begin{picture}(4520.00,4520.00)%
    \gplgaddtomacro\gplbacktext{%
      \colorrgb{0.50,0.50,0.50}
      \put(594,801){\makebox(0,0)[r]{\strut{}$10^{-4}$}}%
      \colorrgb{0.50,0.50,0.50}
      \put(594,1451){\makebox(0,0)[r]{\strut{}$10^{-2}$}}%
      \colorrgb{0.50,0.50,0.50}
      \put(594,2101){\makebox(0,0)[r]{\strut{}$10^{0}$}}%
      \colorrgb{0.50,0.50,0.50}
      \put(594,2752){\makebox(0,0)[r]{\strut{}$10^{2}$}}%
      \colorrgb{0.50,0.50,0.50}
      \put(594,3402){\makebox(0,0)[r]{\strut{}$10^{4}$}}%
      \colorrgb{0.50,0.50,0.50}
      \put(594,4052){\makebox(0,0)[r]{\strut{}$10^{6}$}}%
      \colorrgb{0.50,0.50,0.50}
      \put(730,327){\makebox(0,0){\strut{}$1$}}%
      \colorrgb{0.50,0.50,0.50}
      \put(1282,327){\makebox(0,0){\strut{}$2$}}%
      \colorrgb{0.50,0.50,0.50}
      \put(1833,327){\makebox(0,0){\strut{}$3$}}%
      \colorrgb{0.50,0.50,0.50}
      \put(2385,327){\makebox(0,0){\strut{}$4$}}%
      \colorrgb{0.50,0.50,0.50}
      \put(2936,327){\makebox(0,0){\strut{}$5$}}%
      \colorrgb{0.50,0.50,0.50}
      \put(3487,327){\makebox(0,0){\strut{}$6$}}%
      \colorrgb{0.50,0.50,0.50}
      \put(4039,327){\makebox(0,0){\strut{}$7$}}%
    }%
    \gplgaddtomacro\gplfronttext{%
      \csname LTb\endcsname
      \put(135,2264){\rotatebox{-270}{\makebox(0,0){\strut{}Execution Time $s$}}}%
      \csname LTb\endcsname
      \put(2384,104){\makebox(0,0){\strut{}Number of Qubits $n$}}%
      \csname LTb\endcsname
      \put(2384,4275){\makebox(0,0){\strut{}\texttt{Can-PTM}}}%
    }%
    \gplbacktext
    \put(0,0){\includegraphics[width={226.00bp},height={226.00bp}]{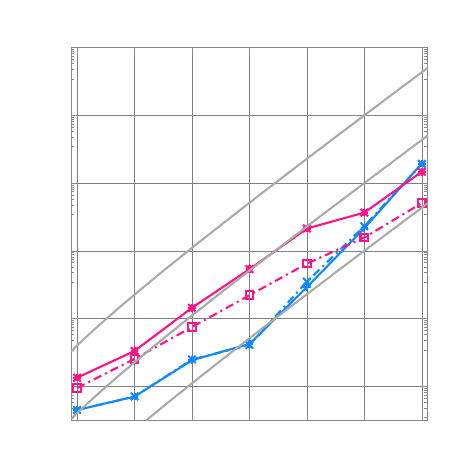}}%
    \gplfronttext
  \end{picture}%
\endgroup

%% file: Kraus-PTM-Tex.tex
\begingroup
  \makeatletter
  \providecommand\color[2][]{%
    \GenericError{(gnuplot) \space\space\space\@spaces}{%
      Package color not loaded in conjunction with
      terminal option `colourtext'%
    }{See the gnuplot documentation for explanation.%
    }{Either use 'blacktext' in gnuplot or load the package
      color.sty in LaTeX.}%
    \renewcommand\color[2][]{}%
  }%
  \providecommand\includegraphics[2][]{%
    \GenericError{(gnuplot) \space\space\space\@spaces}{%
      Package graphicx or graphics not loaded%
    }{See the gnuplot documentation for explanation.%
    }{The gnuplot epslatex terminal needs graphicx.sty or graphics.sty.}%
    \renewcommand\includegraphics[2][]{}%
  }%
  \providecommand\rotatebox[2]{#2}%
  \@ifundefined{ifGPcolor}{%
    \newif\ifGPcolor
    \GPcolortrue
  }{}%
  \@ifundefined{ifGPblacktext}{%
    \newif\ifGPblacktext
    \GPblacktextfalse
  }{}%
  \let\gplgaddtomacro\g@addto@macro
  \gdef\gplbacktext{}%
  \gdef\gplfronttext{}%
  \makeatother
  \ifGPblacktext
    \def\colorrgb#1{}%
    \def\colorgray#1{}%
  \else
    \ifGPcolor
      \def\colorrgb#1{\color[rgb]{#1}}%
      \def\colorgray#1{\color[gray]{#1}}%
      \expandafter\def\csname LTw\endcsname{\color{white}}%
      \expandafter\def\csname LTb\endcsname{\color{black}}%
      \expandafter\def\csname LTa\endcsname{\color{black}}%
      \expandafter\def\csname LT0\endcsname{\color[rgb]{1,0,0}}%
      \expandafter\def\csname LT1\endcsname{\color[rgb]{0,1,0}}%
      \expandafter\def\csname LT2\endcsname{\color[rgb]{0,0,1}}%
      \expandafter\def\csname LT3\endcsname{\color[rgb]{1,0,1}}%
      \expandafter\def\csname LT4\endcsname{\color[rgb]{0,1,1}}%
      \expandafter\def\csname LT5\endcsname{\color[rgb]{1,1,0}}%
      \expandafter\def\csname LT6\endcsname{\color[rgb]{0,0,0}}%
      \expandafter\def\csname LT7\endcsname{\color[rgb]{1,0.3,0}}%
      \expandafter\def\csname LT8\endcsname{\color[rgb]{0.5,0.5,0.5}}%
    \else
      \def\colorrgb#1{\color{black}}%
      \def\colorgray#1{\color[gray]{#1}}%
      \expandafter\def\csname LTw\endcsname{\color{white}}%
      \expandafter\def\csname LTb\endcsname{\color{black}}%
      \expandafter\def\csname LTa\endcsname{\color{black}}%
      \expandafter\def\csname LT0\endcsname{\color{black}}%
      \expandafter\def\csname LT1\endcsname{\color{black}}%
      \expandafter\def\csname LT2\endcsname{\color{black}}%
      \expandafter\def\csname LT3\endcsname{\color{black}}%
      \expandafter\def\csname LT4\endcsname{\color{black}}%
      \expandafter\def\csname LT5\endcsname{\color{black}}%
      \expandafter\def\csname LT6\endcsname{\color{black}}%
      \expandafter\def\csname LT7\endcsname{\color{black}}%
      \expandafter\def\csname LT8\endcsname{\color{black}}%
    \fi
  \fi
    \setlength{\unitlength}{0.0500bp}%
    \ifx\gptboxheight\undefined%
      \newlength{\gptboxheight}%
      \newlength{\gptboxwidth}%
      \newsavebox{\gptboxtext}%
    \fi%
    \setlength{\fboxrule}{0.5pt}%
    \setlength{\fboxsep}{1pt}%
    \definecolor{tbcol}{rgb}{1,1,1}%
\begin{picture}(4520.00,4520.00)%
    \gplgaddtomacro\gplbacktext{%
      \colorrgb{0.50,0.50,0.50}
      \put(594,801){\makebox(0,0)[r]{\strut{}$10^{-4}$}}%
      \colorrgb{0.50,0.50,0.50}
      \put(594,1451){\makebox(0,0)[r]{\strut{}$10^{-2}$}}%
      \colorrgb{0.50,0.50,0.50}
      \put(594,2101){\makebox(0,0)[r]{\strut{}$10^{0}$}}%
      \colorrgb{0.50,0.50,0.50}
      \put(594,2752){\makebox(0,0)[r]{\strut{}$10^{2}$}}%
      \colorrgb{0.50,0.50,0.50}
      \put(594,3402){\makebox(0,0)[r]{\strut{}$10^{4}$}}%
      \colorrgb{0.50,0.50,0.50}
      \put(594,4052){\makebox(0,0)[r]{\strut{}$10^{6}$}}%
      \colorrgb{0.50,0.50,0.50}
      \put(730,327){\makebox(0,0){\strut{}$1$}}%
      \colorrgb{0.50,0.50,0.50}
      \put(1282,327){\makebox(0,0){\strut{}$2$}}%
      \colorrgb{0.50,0.50,0.50}
      \put(1833,327){\makebox(0,0){\strut{}$3$}}%
      \colorrgb{0.50,0.50,0.50}
      \put(2385,327){\makebox(0,0){\strut{}$4$}}%
      \colorrgb{0.50,0.50,0.50}
      \put(2936,327){\makebox(0,0){\strut{}$5$}}%
      \colorrgb{0.50,0.50,0.50}
      \put(3487,327){\makebox(0,0){\strut{}$6$}}%
      \colorrgb{0.50,0.50,0.50}
      \put(4039,327){\makebox(0,0){\strut{}$7$}}%
    }%
    \gplgaddtomacro\gplfronttext{%
      \csname LTb\endcsname
      \put(135,2264){\rotatebox{-270}{\makebox(0,0){\strut{}Execution Time $s$}}}%
      \csname LTb\endcsname
      \put(2384,104){\makebox(0,0){\strut{}Number of Qubits $n$}}%
      \csname LTb\endcsname
      \put(2384,4275){\makebox(0,0){\strut{}\texttt{Kraus-PTM}}}%
    }%
    \gplbacktext
    \put(0,0){\includegraphics[width={226.00bp},height={226.00bp}]{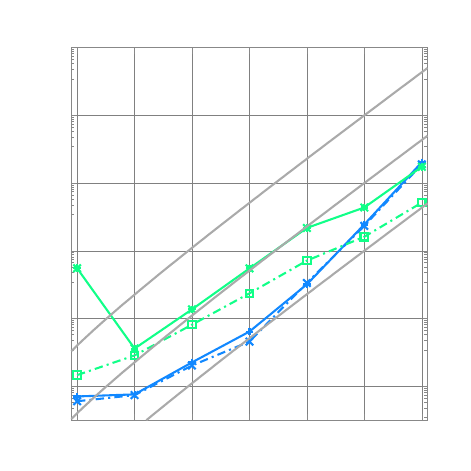}}%
    \gplfronttext
  \end{picture}%
\endgroup

%% file: Choi-PTM-Tex.tex
\begingroup
  \makeatletter
  \providecommand\color[2][]{%
    \GenericError{(gnuplot) \space\space\space\@spaces}{%
      Package color not loaded in conjunction with
      terminal option `colourtext'%
    }{See the gnuplot documentation for explanation.%
    }{Either use 'blacktext' in gnuplot or load the package
      color.sty in LaTeX.}%
    \renewcommand\color[2][]{}%
  }%
  \providecommand\includegraphics[2][]{%
    \GenericError{(gnuplot) \space\space\space\@spaces}{%
      Package graphicx or graphics not loaded%
    }{See the gnuplot documentation for explanation.%
    }{The gnuplot epslatex terminal needs graphicx.sty or graphics.sty.}%
    \renewcommand\includegraphics[2][]{}%
  }%
  \providecommand\rotatebox[2]{#2}%
  \@ifundefined{ifGPcolor}{%
    \newif\ifGPcolor
    \GPcolortrue
  }{}%
  \@ifundefined{ifGPblacktext}{%
    \newif\ifGPblacktext
    \GPblacktextfalse
  }{}%
  \let\gplgaddtomacro\g@addto@macro
  \gdef\gplbacktext{}%
  \gdef\gplfronttext{}%
  \makeatother
  \ifGPblacktext
    \def\colorrgb#1{}%
    \def\colorgray#1{}%
  \else
    \ifGPcolor
      \def\colorrgb#1{\color[rgb]{#1}}%
      \def\colorgray#1{\color[gray]{#1}}%
      \expandafter\def\csname LTw\endcsname{\color{white}}%
      \expandafter\def\csname LTb\endcsname{\color{black}}%
      \expandafter\def\csname LTa\endcsname{\color{black}}%
      \expandafter\def\csname LT0\endcsname{\color[rgb]{1,0,0}}%
      \expandafter\def\csname LT1\endcsname{\color[rgb]{0,1,0}}%
      \expandafter\def\csname LT2\endcsname{\color[rgb]{0,0,1}}%
      \expandafter\def\csname LT3\endcsname{\color[rgb]{1,0,1}}%
      \expandafter\def\csname LT4\endcsname{\color[rgb]{0,1,1}}%
      \expandafter\def\csname LT5\endcsname{\color[rgb]{1,1,0}}%
      \expandafter\def\csname LT6\endcsname{\color[rgb]{0,0,0}}%
      \expandafter\def\csname LT7\endcsname{\color[rgb]{1,0.3,0}}%
      \expandafter\def\csname LT8\endcsname{\color[rgb]{0.5,0.5,0.5}}%
    \else
      \def\colorrgb#1{\color{black}}%
      \def\colorgray#1{\color[gray]{#1}}%
      \expandafter\def\csname LTw\endcsname{\color{white}}%
      \expandafter\def\csname LTb\endcsname{\color{black}}%
      \expandafter\def\csname LTa\endcsname{\color{black}}%
      \expandafter\def\csname LT0\endcsname{\color{black}}%
      \expandafter\def\csname LT1\endcsname{\color{black}}%
      \expandafter\def\csname LT2\endcsname{\color{black}}%
      \expandafter\def\csname LT3\endcsname{\color{black}}%
      \expandafter\def\csname LT4\endcsname{\color{black}}%
      \expandafter\def\csname LT5\endcsname{\color{black}}%
      \expandafter\def\csname LT6\endcsname{\color{black}}%
      \expandafter\def\csname LT7\endcsname{\color{black}}%
      \expandafter\def\csname LT8\endcsname{\color{black}}%
    \fi
  \fi
    \setlength{\unitlength}{0.0500bp}%
    \ifx\gptboxheight\undefined%
      \newlength{\gptboxheight}%
      \newlength{\gptboxwidth}%
      \newsavebox{\gptboxtext}%
    \fi%
    \setlength{\fboxrule}{0.5pt}%
    \setlength{\fboxsep}{1pt}%
    \definecolor{tbcol}{rgb}{1,1,1}%
\begin{picture}(4520.00,4520.00)%
    \gplgaddtomacro\gplbacktext{%
      \colorrgb{0.50,0.50,0.50}
      \put(594,801){\makebox(0,0)[r]{\strut{}$10^{-4}$}}%
      \colorrgb{0.50,0.50,0.50}
      \put(594,1451){\makebox(0,0)[r]{\strut{}$10^{-2}$}}%
      \colorrgb{0.50,0.50,0.50}
      \put(594,2101){\makebox(0,0)[r]{\strut{}$10^{0}$}}%
      \colorrgb{0.50,0.50,0.50}
      \put(594,2752){\makebox(0,0)[r]{\strut{}$10^{2}$}}%
      \colorrgb{0.50,0.50,0.50}
      \put(594,3402){\makebox(0,0)[r]{\strut{}$10^{4}$}}%
      \colorrgb{0.50,0.50,0.50}
      \put(594,4052){\makebox(0,0)[r]{\strut{}$10^{6}$}}%
      \colorrgb{0.50,0.50,0.50}
      \put(730,327){\makebox(0,0){\strut{}$1$}}%
      \colorrgb{0.50,0.50,0.50}
      \put(1282,327){\makebox(0,0){\strut{}$2$}}%
      \colorrgb{0.50,0.50,0.50}
      \put(1833,327){\makebox(0,0){\strut{}$3$}}%
      \colorrgb{0.50,0.50,0.50}
      \put(2385,327){\makebox(0,0){\strut{}$4$}}%
      \colorrgb{0.50,0.50,0.50}
      \put(2936,327){\makebox(0,0){\strut{}$5$}}%
      \colorrgb{0.50,0.50,0.50}
      \put(3487,327){\makebox(0,0){\strut{}$6$}}%
      \colorrgb{0.50,0.50,0.50}
      \put(4039,327){\makebox(0,0){\strut{}$7$}}%
    }%
    \gplgaddtomacro\gplfronttext{%
      \csname LTb\endcsname
      \put(135,2264){\rotatebox{-270}{\makebox(0,0){\strut{}Execution Time $s$}}}%
      \csname LTb\endcsname
      \put(2384,104){\makebox(0,0){\strut{}Number of Qubits $n$}}%
      \csname LTb\endcsname
      \put(2384,4275){\makebox(0,0){\strut{}\texttt{Choi-PTM}}}%
    }%
    \gplbacktext
    \put(0,0){\includegraphics[width={226.00bp},height={226.00bp}]{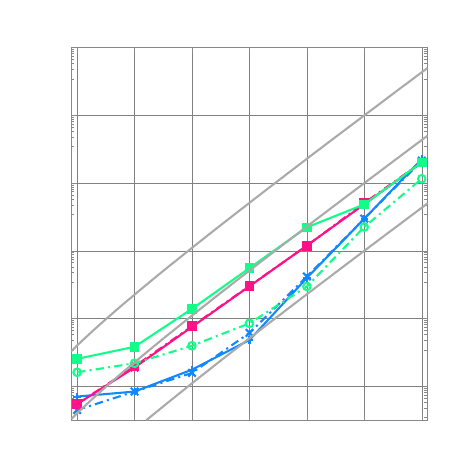}}%
    \gplfronttext
  \end{picture}%
\endgroup

%% file: Chi-PTM-Tex.tex
\begingroup
  \makeatletter
  \providecommand\color[2][]{%
    \GenericError{(gnuplot) \space\space\space\@spaces}{%
      Package color not loaded in conjunction with
      terminal option `colourtext'%
    }{See the gnuplot documentation for explanation.%
    }{Either use 'blacktext' in gnuplot or load the package
      color.sty in LaTeX.}%
    \renewcommand\color[2][]{}%
  }%
  \providecommand\includegraphics[2][]{%
    \GenericError{(gnuplot) \space\space\space\@spaces}{%
      Package graphicx or graphics not loaded%
    }{See the gnuplot documentation for explanation.%
    }{The gnuplot epslatex terminal needs graphicx.sty or graphics.sty.}%
    \renewcommand\includegraphics[2][]{}%
  }%
  \providecommand\rotatebox[2]{#2}%
  \@ifundefined{ifGPcolor}{%
    \newif\ifGPcolor
    \GPcolortrue
  }{}%
  \@ifundefined{ifGPblacktext}{%
    \newif\ifGPblacktext
    \GPblacktextfalse
  }{}%
  \let\gplgaddtomacro\g@addto@macro
  \gdef\gplbacktext{}%
  \gdef\gplfronttext{}%
  \makeatother
  \ifGPblacktext
    \def\colorrgb#1{}%
    \def\colorgray#1{}%
  \else
    \ifGPcolor
      \def\colorrgb#1{\color[rgb]{#1}}%
      \def\colorgray#1{\color[gray]{#1}}%
      \expandafter\def\csname LTw\endcsname{\color{white}}%
      \expandafter\def\csname LTb\endcsname{\color{black}}%
      \expandafter\def\csname LTa\endcsname{\color{black}}%
      \expandafter\def\csname LT0\endcsname{\color[rgb]{1,0,0}}%
      \expandafter\def\csname LT1\endcsname{\color[rgb]{0,1,0}}%
      \expandafter\def\csname LT2\endcsname{\color[rgb]{0,0,1}}%
      \expandafter\def\csname LT3\endcsname{\color[rgb]{1,0,1}}%
      \expandafter\def\csname LT4\endcsname{\color[rgb]{0,1,1}}%
      \expandafter\def\csname LT5\endcsname{\color[rgb]{1,1,0}}%
      \expandafter\def\csname LT6\endcsname{\color[rgb]{0,0,0}}%
      \expandafter\def\csname LT7\endcsname{\color[rgb]{1,0.3,0}}%
      \expandafter\def\csname LT8\endcsname{\color[rgb]{0.5,0.5,0.5}}%
    \else
      \def\colorrgb#1{\color{black}}%
      \def\colorgray#1{\color[gray]{#1}}%
      \expandafter\def\csname LTw\endcsname{\color{white}}%
      \expandafter\def\csname LTb\endcsname{\color{black}}%
      \expandafter\def\csname LTa\endcsname{\color{black}}%
      \expandafter\def\csname LT0\endcsname{\color{black}}%
      \expandafter\def\csname LT1\endcsname{\color{black}}%
      \expandafter\def\csname LT2\endcsname{\color{black}}%
      \expandafter\def\csname LT3\endcsname{\color{black}}%
      \expandafter\def\csname LT4\endcsname{\color{black}}%
      \expandafter\def\csname LT5\endcsname{\color{black}}%
      \expandafter\def\csname LT6\endcsname{\color{black}}%
      \expandafter\def\csname LT7\endcsname{\color{black}}%
      \expandafter\def\csname LT8\endcsname{\color{black}}%
    \fi
  \fi
    \setlength{\unitlength}{0.0500bp}%
    \ifx\gptboxheight\undefined%
      \newlength{\gptboxheight}%
      \newlength{\gptboxwidth}%
      \newsavebox{\gptboxtext}%
    \fi%
    \setlength{\fboxrule}{0.5pt}%
    \setlength{\fboxsep}{1pt}%
    \definecolor{tbcol}{rgb}{1,1,1}%
\begin{picture}(4520.00,4520.00)%
    \gplgaddtomacro\gplbacktext{%
      \colorrgb{0.50,0.50,0.50}
      \put(594,801){\makebox(0,0)[r]{\strut{}$10^{-4}$}}%
      \colorrgb{0.50,0.50,0.50}
      \put(594,1451){\makebox(0,0)[r]{\strut{}$10^{-2}$}}%
      \colorrgb{0.50,0.50,0.50}
      \put(594,2101){\makebox(0,0)[r]{\strut{}$10^{0}$}}%
      \colorrgb{0.50,0.50,0.50}
      \put(594,2752){\makebox(0,0)[r]{\strut{}$10^{2}$}}%
      \colorrgb{0.50,0.50,0.50}
      \put(594,3402){\makebox(0,0)[r]{\strut{}$10^{4}$}}%
      \colorrgb{0.50,0.50,0.50}
      \put(594,4052){\makebox(0,0)[r]{\strut{}$10^{6}$}}%
      \colorrgb{0.50,0.50,0.50}
      \put(730,327){\makebox(0,0){\strut{}$1$}}%
      \colorrgb{0.50,0.50,0.50}
      \put(1282,327){\makebox(0,0){\strut{}$2$}}%
      \colorrgb{0.50,0.50,0.50}
      \put(1833,327){\makebox(0,0){\strut{}$3$}}%
      \colorrgb{0.50,0.50,0.50}
      \put(2385,327){\makebox(0,0){\strut{}$4$}}%
      \colorrgb{0.50,0.50,0.50}
      \put(2936,327){\makebox(0,0){\strut{}$5$}}%
      \colorrgb{0.50,0.50,0.50}
      \put(3487,327){\makebox(0,0){\strut{}$6$}}%
      \colorrgb{0.50,0.50,0.50}
      \put(4039,327){\makebox(0,0){\strut{}$7$}}%
    }%
    \gplgaddtomacro\gplfronttext{%
      \csname LTb\endcsname
      \put(135,2264){\rotatebox{-270}{\makebox(0,0){\strut{}Execution Time $s$}}}%
      \csname LTb\endcsname
      \put(2384,104){\makebox(0,0){\strut{}Number of Qubits $n$}}%
      \csname LTb\endcsname
      \put(2384,4275){\makebox(0,0){\strut{}\texttt{Chi-PTM}}}%
    }%
    \gplbacktext
    \put(0,0){\includegraphics[width={226.00bp},height={226.00bp}]{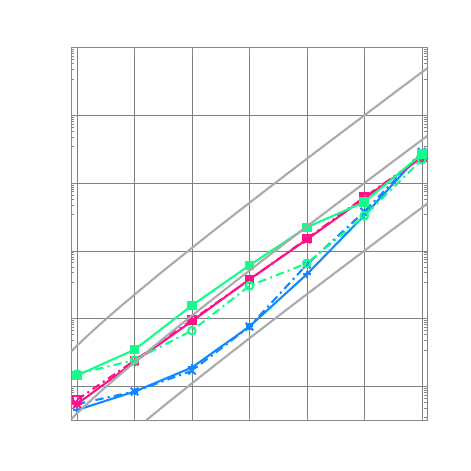}}%
    \gplfronttext
  \end{picture}%
\endgroup